\documentclass[useAMS,usenatbib]{mn2e}

\usepackage{natbib}
\usepackage{graphicx}
\usepackage{txfonts}
\usepackage{url}
\usepackage{times}
\usepackage{pdflscape}

\def\aj{AJ}%
\def\araa{ARA\&A}%
\def\apj{ApJ}%
\def\apjl{ApJ}%
\def\apss{Ap\&SS}%
\def\aap{A\&A}%
\def\mnras{MNRAS}%
\def\prd{Phys.~Rev.~D}%
\def\pasp{PASP}%
\def\nat{Nature}%
\def\ms{\,m\,s$^{-1}$}         
\def\kms{\,km\,s$^{-1}$}       
\def\msol{$M_\odot$}		
\def\rsol{$R_\odot$}		
\def\denssol{$\rho_\odot$}	
\def\rstar{$R_*$}		
\def\mplanet{$M_{\rm P}$}	
\def\rplanet{$R_{\rm P}$}	
\def\mjup{$M_{\rm Jup}$}	
\def\rjup{$R_{\rm Jup}$}	
\def\teql{$T_{\rm eql}$}
\def\teff{$T_{\rm eff}$}
\def\feh{[Fe/H]}
\def\logg{$\log g_*$}
\def\vsini{$v \sin i_*$}
\def\mictrb{$\xi_{\rm t}$}

\def\kms{km\, s$^{-1}$}

\def\svsicos{$\sqrt{v \sin I} \cos \lambda$}
\def\svsisin{$\sqrt{v \sin I} \sin \lambda$}
\def\chisq{$\chi^2$}

\def\prot{$P_{\rm rot}$}
\def\porb{$P_{\rm orb}$}

\newcommand{\Mstar}{\mbox{${M}_{\star}$}}
\newcommand{\rhostar}{\mbox{${\rho}_{\star}$}}
\newcommand{\taustar}{\mbox{${\tau}_{\star}$}}

\newcommand{\Teff}{\mbox{T$_{\rmn{eff}}$}}
\newcommand{\amlt}{\mbox{$\alpha_{\rmn{MLT}}$}}
\newcommand{\leftcell}[1]{\multicolumn{1}{l}{#1}}

\title[Six hot Jupiters transiting F/G stars]{Six newly-discovered hot 
Jupiters transiting F/G stars: WASP-87b, WASP-108b, WASP-109b, WASP-110b, 
WASP-111b \& WASP-112b\thanks{Based on 
observations made with the WASP-South photometric survey instrument at SAAO and, all located at La Silla: 
the 1.54-m Danish photometer under progam CN2013A-159; the 60-cm TRAPPIST photometer; and the EulerCam photometer 
and the CORALIE spectrograph, both mounted on the 1.2-m Euler-Swiss telescope.
}}

\author[D.~R.~Anderson et al.]{D.~R.~Anderson,$^{1}$\thanks{E-mail: d.r.anderson@keele.ac.uk}
D.~J.~A.~Brown,$^{2}$
A.~Collier~Cameron,$^{3}$
L.~Delrez,$^{4}$ 
A.~Fumel,$^{4}$ 
M.~Gillon,$^{4}$ 
\newauthor
C.~Hellier,$^{1}$ 
E.~Jehin,$^{4}$ 
M.~Lendl,$^{4,5}$ 
P.~F.~L.~Maxted,$^{1}$ 
M.~Neveu-VanMalle,$^{5,6}$ 
F.~Pepe,$^{4}$ 
\newauthor
D.~Pollacco,$^{2}$ 
D.~Queloz,$^{5,6}$ 
P.~Rojo,$^{7}$ 
D.~S\'egransan,$^{5}$ 
A.~M.~Serenelli,$^{8}$ 
B.~Smalley,$^{1}$ 
\newauthor
A.~M.~S.~Smith,$^{1,9}$ 
J.~Southworth,$^{1}$ 
A.~H.~M.~J.~Triaud,$^{5,10}$ 
O.~D.~Turner,$^{1}$ 
S.~Udry,$^{5}$ and 
\newauthor
R.~G.~West$^{2}$\\
$^1$Astrophysics Group, Keele University, Staffordshire ST5 5BG, UK\\
$^2$Department of Physics, University of Warwick, Coventry CV4 7AL, UK\\
$^3$SUPA, School of Physics and Astronomy, University of St. Andrews, 
           North Haugh, Fife KY16 9SS, UK\\
$^4$Institut d'Astrophysique et de G\'eophysique,  Universit\'e de 
           Li\`ege,  All\'ee du 6 Ao\^ut, 17,  Bat.  B5C, Li\`ege 1, Belgium\\
$^5$Observatoire de Gen\`eve, Universit\'e de Gen\`eve, 51 Chemin 
           des Maillettes, 1290 Sauverny, Switzerland\\
$^6$Cavendish Laboratory, J J Thomson Avenue, Cambridge, CB3 0HE, UK\\
$^7$Departamento de Astronomia, Universidad de Chile, Santiago, Chile\\
$^8$Instituto de Ciencias del Espacio (CSIC-IEEC), Facultad de Ciencias, 
           Campus UAB, 08193, Bellaterra, Spain\\
$^9$N. Copernicus Astronomical Centre, Polish Academy of Sciences, 
           Bartycka 18, 00-716, Warsaw, Poland\\
$^{10}$Department of Physics, Massachusetts Institute of Technology, Cambridge, 
           MA 02139, USA\\
}

\begin{document}

\date{Accepted Year Month Day. Received Year Month Day; in original form Year Month Day}

\pagerange{\pageref{firstpage}--\pageref{lastpage}} \pubyear{2014}

\maketitle

\label{firstpage}

\begin{abstract}
We present the discoveries of six transiting hot Jupiters: WASP-87b, WASP-108b, WASP-109b, WASP-110b, WASP-111b and WASP-112b. 
The planets have masses of 0.51--2.2\,\mjup\ and radii of 1.19--1.44\,\rjup\ and are in orbits of 1.68--3.78\,d around stars with masses 0.81--1.50\,\msol.

WASP-111b is in a prograde, near-aligned ($\lambda = -5 \pm 16^\circ$), near-circular ($e < 0.10$ at 2\,$\sigma$) orbit around a mid-F star. 
As tidal alignment around such a hot star is thought to be inefficient, this suggests that either the planet 
migrated inwards through the protoplanetary disc or that scattering processes happened to leave it in a near-aligned orbit.
WASP-111 appears to have transitioned from an active to a quiescent state between the 2012 and 2013 seasons, 
which makes the system a candidate for studying the effects of variable activity on a hot-Jupiter atmosphere. 
We find evidence that the mid-F star WASP-87 is a visual binary with a mid-G star. 
Two host stars are metal poor: WASP-112 has \feh\ = $-0.64 \pm 0.15$ and WASP-87 has \feh\ = $-0.41 \pm 0.10$.
The low density of WASP-112 (0.81\,\msol, $0.80 \pm 0.04$\,\denssol) cannot be matched by standard models for any reasonable value of the age of the star, suggesting it to be affected by the ``radius anomaly''. 
\end{abstract}

\begin{keywords}
planets and satellites: individual: WASP-87b -- 
planets and satellites: individual: WASP-108b -- 
planets and satellites: individual: WASP-109b -- 
planets and satellites: individual: WASP-110b -- 
planets and satellites: individual: WASP-111b -- 
planets and satellites: individual: WASP-112b.
\end{keywords}

\section{Introduction}
Advances are made in the understanding of planet formation and evolution by studying large samples of exoplanet systems and smaller subsets of well-characterised systems. 
There is a preponderance of giant planets around metal-rich stars and a dearth around metal-poor stars, 
suggestive of the core-accretion model of planet formation \citep{1997MNRAS.285..403G, 2004A&A...415.1153S, 2005ApJ...622.1102F}. 
In a study of candidate transiting planets found by {\it Kepler} \citep{2010Sci...327..977B}, \citet{2013ApJ...775L..11M} noted a dearth of planets in orbits shorter than 2--3\,d around stars with rotation periods shorter than 5--10\,d, in which tidal interactions seem likely to have played a role \citep{2014MNRAS.443.1451L, 2014ApJ...786..139T, 2014ApJ...787..131Z}.

With coverage of bright stars ($V$ = 9--13) across the whole sky (other than the galactic plane and the celestial poles) the SuperWASP photometric survey is discovering some of those inherently rare systems from which we stand, perhaps, to learn the most \citep{2006PASP..118.1407P}. 
For example: WASP-58 with \feh\ = $-$0.45 and WASP-98 with \feh\ = $-$0.60 are hot Jupiters orbiting metal-poor stars \citep{2013A&A...549A.134H, 2014MNRAS.440.1982H}; and WASP-18b (10.5\,\mjup), WASP-33b (3.3\,\mjup) and WASP-103b (1.49\,\mjup) are in the gap noted by \citet{2013ApJ...775L..11M}: they are in $\sim$1-d orbits around rapidly rotating, bright stars ($P_{\rm rot} < 7$\,d; $V$ = 8.3--12.5; \citealt{2009Natur.460.1098H, 2010MNRAS.407..507C, 2014A&A...562L...3G}). 

In this paper, we present the discoveries of six transiting hot Jupiters, two of which orbit metal-poor stars and three of which are in short orbits around fast rotators. 

\section{Observations}
\label{sec:obs}
WASP-South images one third of the visible South-African sky (avoiding the galactic plane and the south pole) every $\sim$10 minutes and is sensitive to the detection of giant planets transiting bright stars ($V$ = 9--13). 
The survey and the search techniques are described in 
\citet{2006PASP..118.1407P} and \citet{2006MNRAS.373..799C, 2007MNRAS.380.1230C} and the analysis techniques are described in detail in recent WASP discovery papers (e.g. \citealt{2014MNRAS.445.1114A}).

We routinely investigate the promising transit signals that we find in WASP lightcurves photometrically with the 0.6-m TRAPPIST robotic photometer and the EulerCam photometer and spectroscopically with the CORALIE spectrograph \citep{2011A&A...533A..88G, 2012A&A...544A..72L, 2000A&A...354...99Q}; both EulerCam and CORALIE are mounted on the 1.2-m Euler-Swiss telescope. 
For three of the systems presented herein we obtained additional transit photometry with the 1.54-m Danish telescope. 
All the follow-up instruments are situated at La Silla. 

\begin{figure}
\includegraphics[width=90mm]{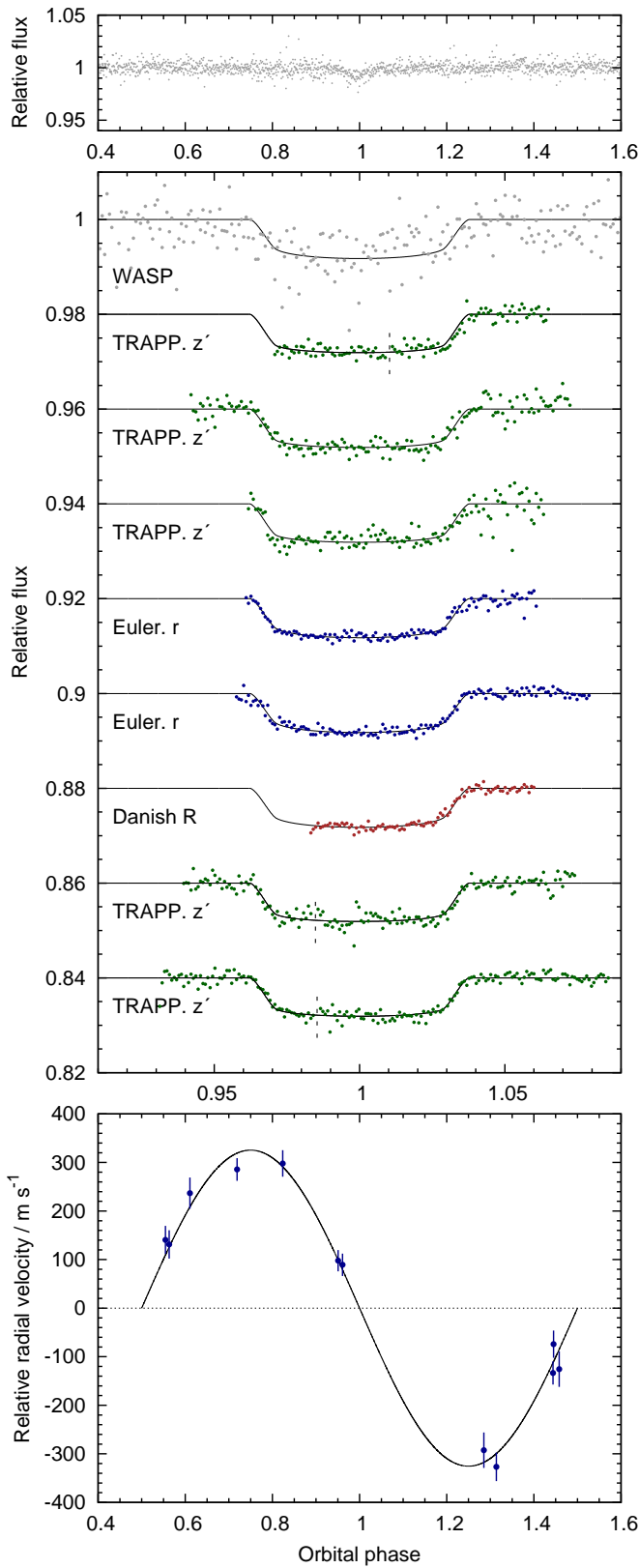}
\caption{WASP-87b discovery data. 
Top panel: WASP lightcurve folded on the transit ephemeris. 
Second panel: Transit lightcurves from facilities and in passbands as labelled, offset for clarity. 
Each photometry set was binned in phase with a bin width equivalent to two minutes. 
The best-fitting transit model is superimposed. Vertical dashed lines indicate partitioning due to 
meridian flips or other issues. 
Third panel: The CORALIE radial velocities with the best-fitting circular Keplerian orbit model. 
\label{fig:w87-rv-phot}} 
\end{figure}

\begin{figure}
\includegraphics[width=90mm]{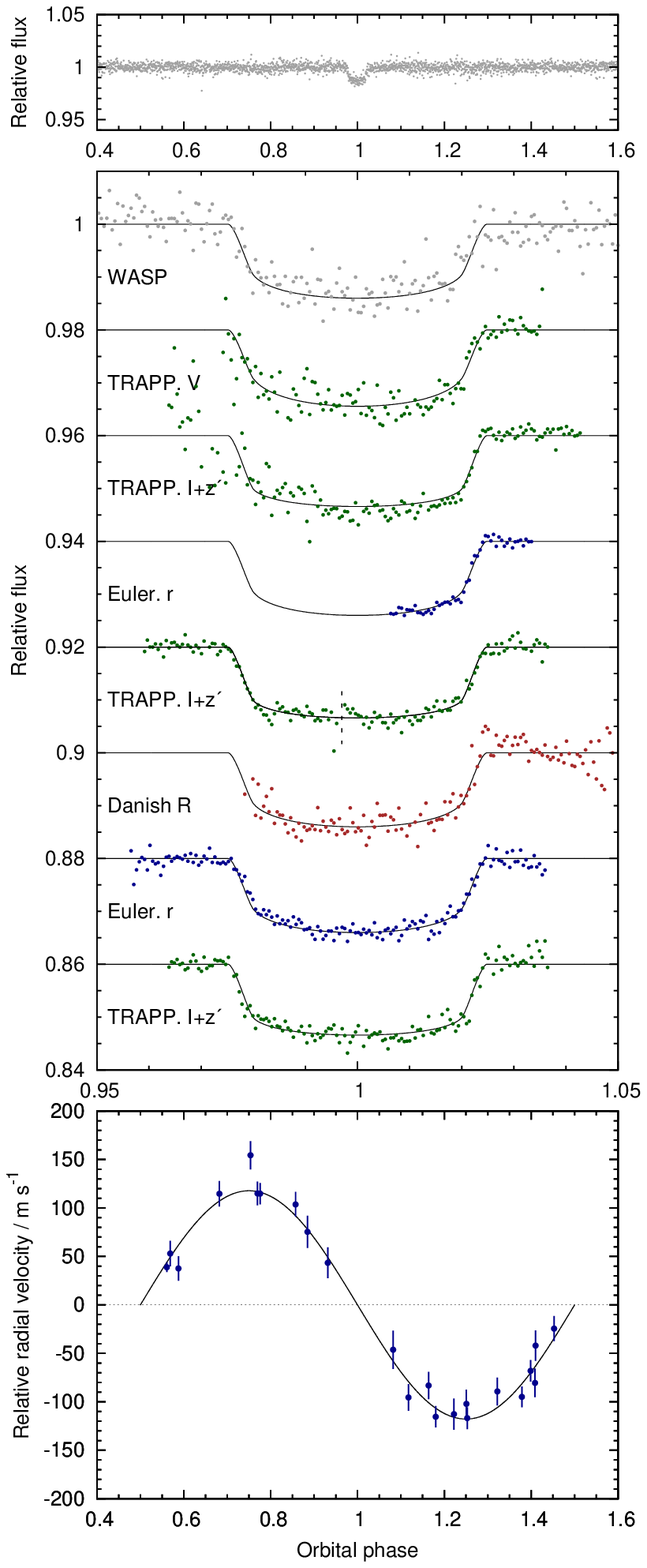}
\caption{WASP-108b discovery data. 
Caption as for Fig.~\ref{fig:w87-rv-phot}. 
\label{fig:w108-rv-phot}} 
\end{figure}

\begin{figure}
\includegraphics[width=90mm]{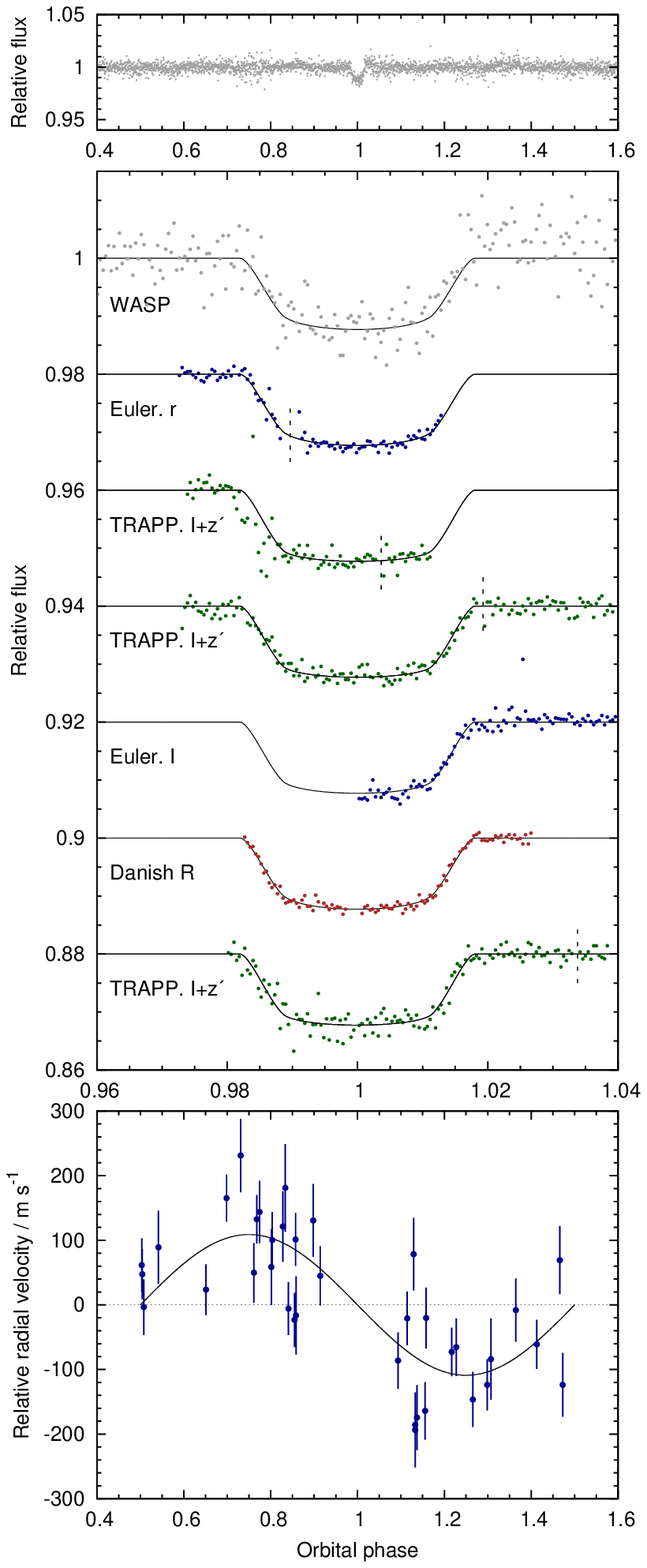}
\caption{WASP-109b discovery data. 
Caption as for Fig.~\ref{fig:w87-rv-phot}. 
\label{fig:w109-rv-phot}} 
\end{figure}

\begin{figure}
\includegraphics[width=90mm]{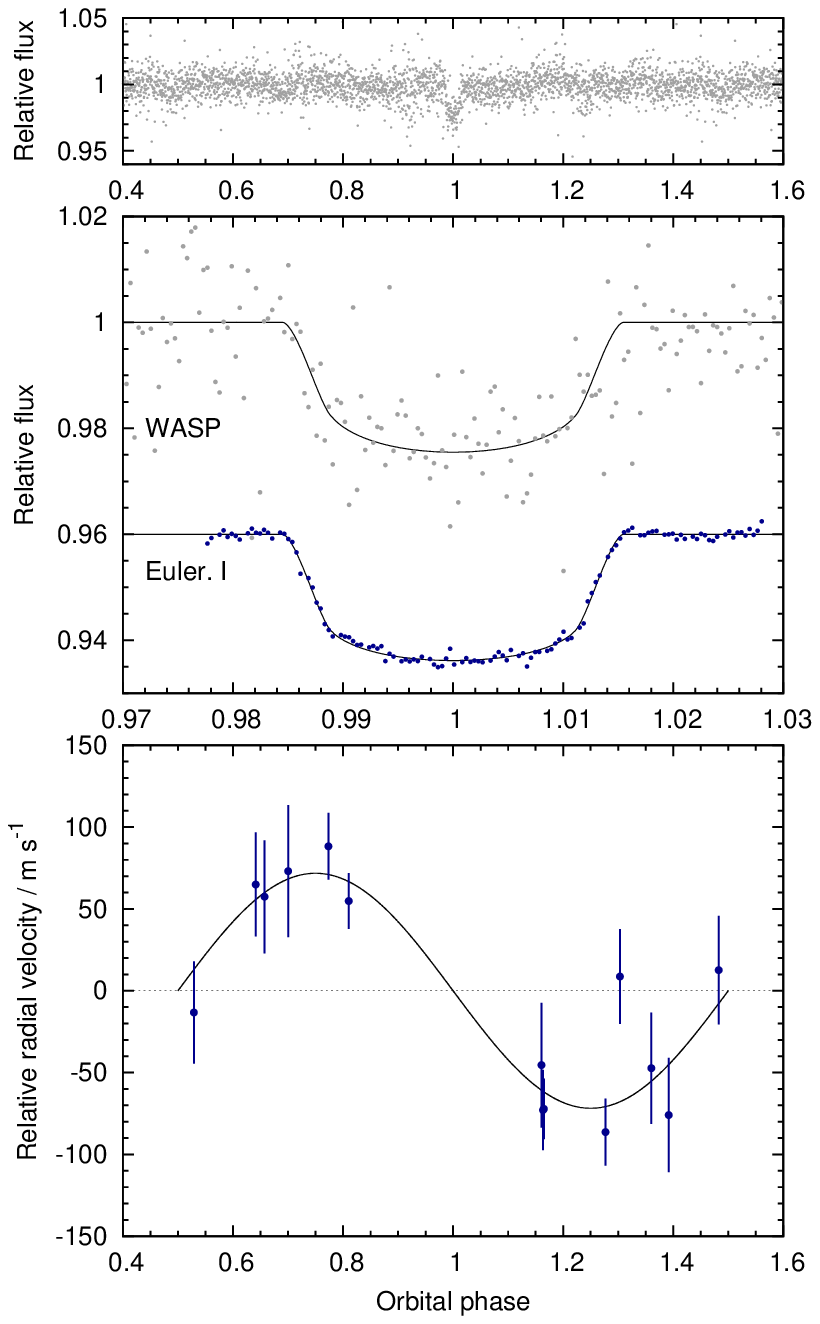}
\caption{WASP-110b discovery data. 
Caption as for Fig.~\ref{fig:w87-rv-phot}. 
\label{fig:w110-rv-phot}} 
\end{figure}

\begin{figure}
\includegraphics[width=90mm]{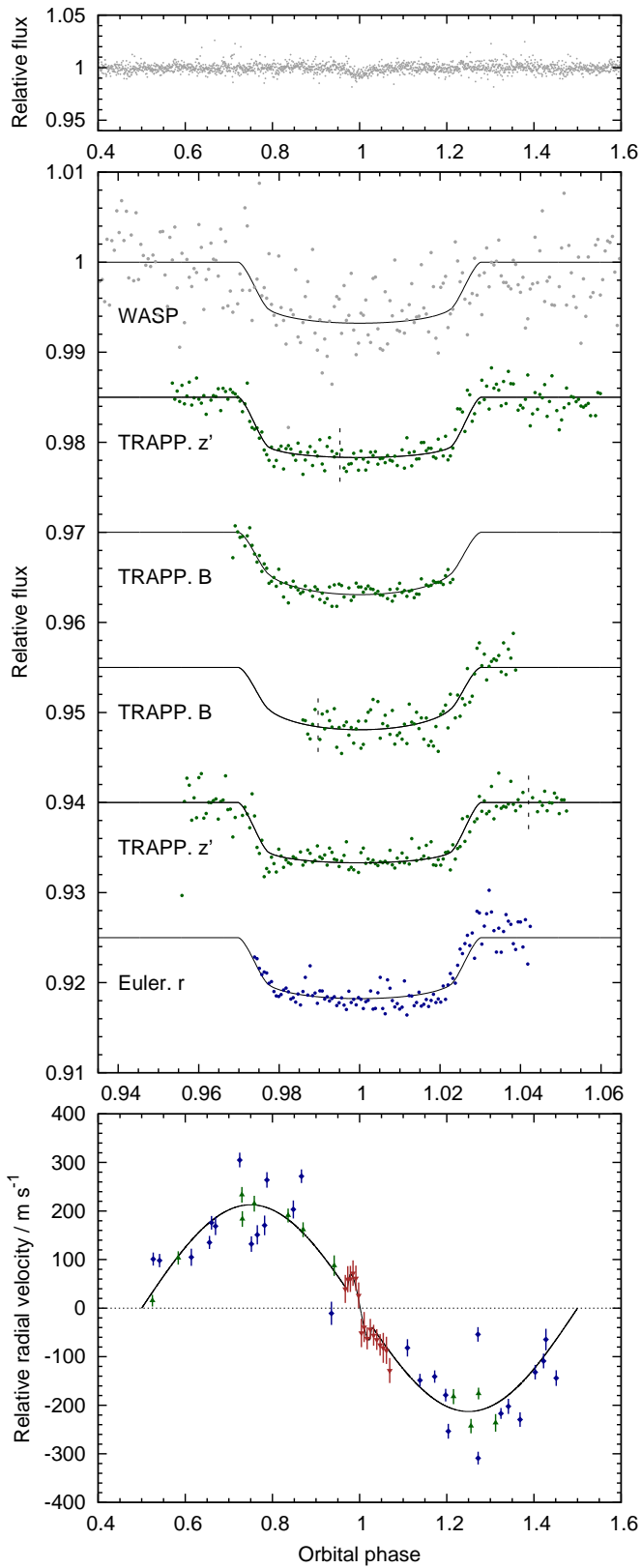}
\caption{WASP-111b discovery data. 
Caption as for Fig.~\ref{fig:w87-rv-phot}. 
The RVs from 2012, 2013 and the RM observation are plotted, respectively, as blue diamonds, green down-triangles 
and brown up-triangles.
\label{fig:w111-rv-phot}} 
\end{figure}

\begin{figure}
\includegraphics[width=90mm]{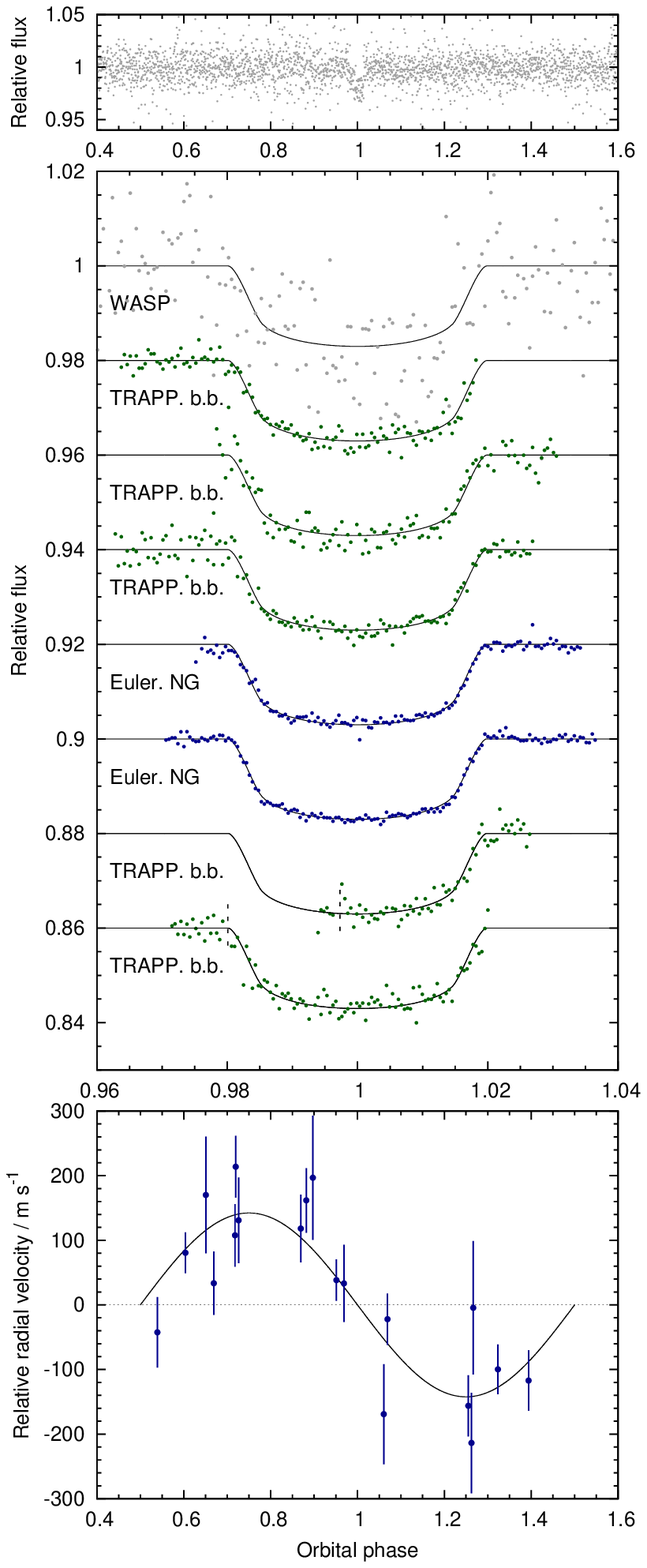}
\caption{WASP-112b discovery data. 
Caption as for Fig.~\ref{fig:w87-rv-phot}. 
\label{fig:w112-rv-phot}} 
\end{figure}

\begin{figure*}
\includegraphics[]{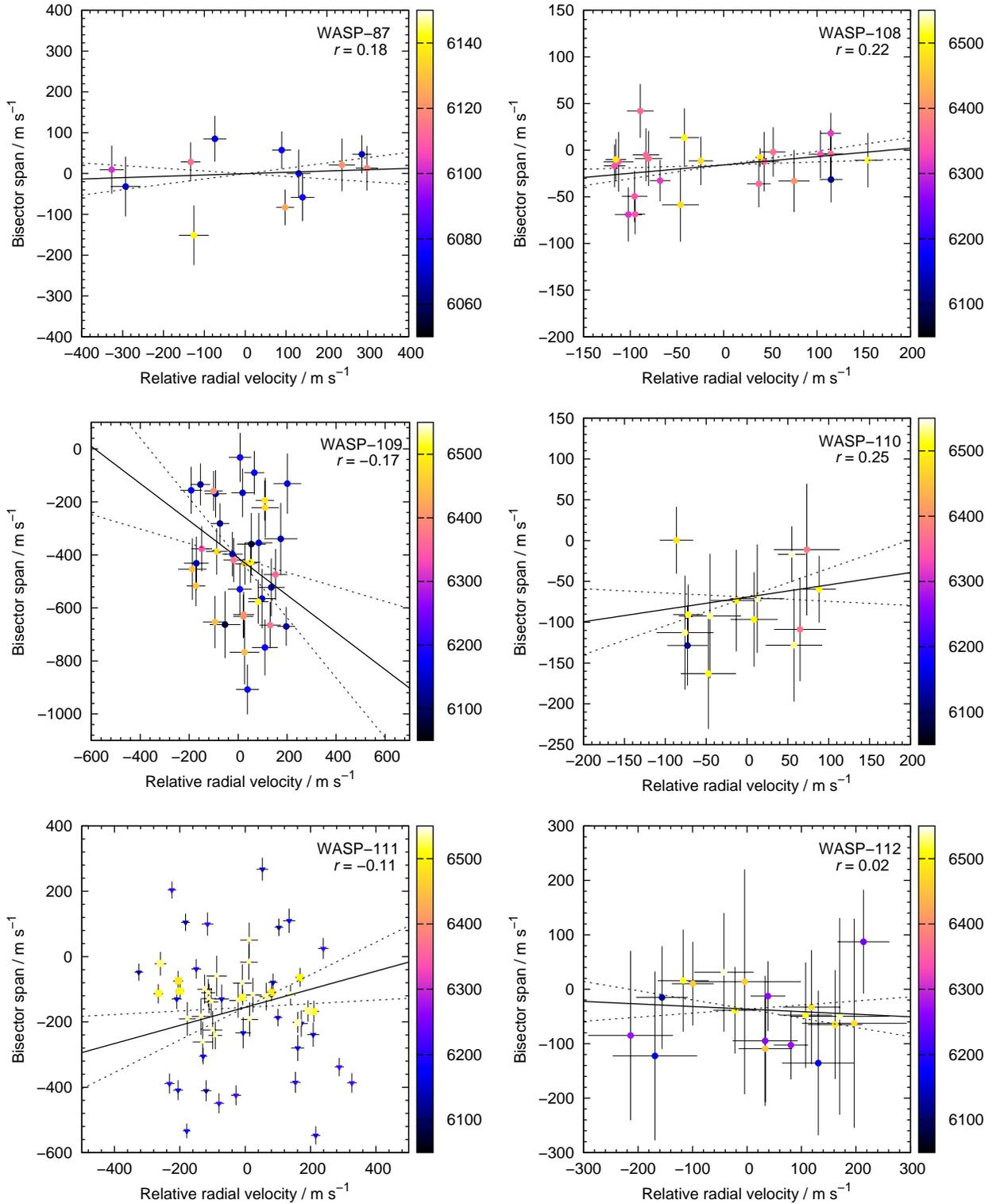}
\caption{The lack of correlation between bisector span and radial velocity for the six stars excludes transit mimics. 
The solid line is the best linear fit to the data and the dotted lines are the 1-$\sigma$ limits on the gradient. 
The Pearson product-moment correlation coefficient, $r$, is given in each panel. 
The Julian date of the observation (BJD $-$ 2\,450\,000) is represented by the symbol colour.
\label{fig:bis}} 
\end{figure*}

The radial-velocity (RV) measurements computed from the CORALIE spectra exhibit variations with similar periods as the photometric dimmings seen in the WASP lightcurves and with amplitudes consistent with planetary-mass companions. 
The photometry and RVs are plotted for each system in Figures~\ref{fig:w87-rv-phot}, \ref{fig:w108-rv-phot}, \ref{fig:w109-rv-phot}, \ref{fig:w110-rv-phot}, \ref{fig:w111-rv-phot} and \ref{fig:w112-rv-phot}.
The lack of any significant correlation between bisector span and RV supports our conclusion that the observed periodic dimmings and RV variations are caused by transiting planets (Figure~\ref{fig:bis}). 

\begin{figure}
\includegraphics[width=90mm]{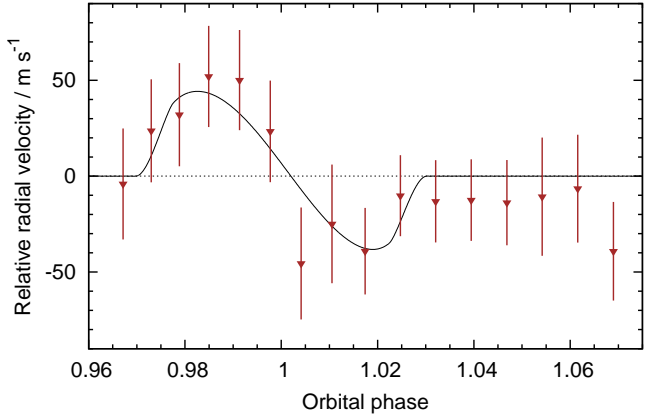}
\caption{The Rossiter-McLaughlin effect of WASP-111b. 
\label{fig:w111-rm}} 
\end{figure}

With the aim of measuring the Rossiter-McLaughlin (RM) effect and thus determining the degree of alignment between the stars's rotation and the planet's orbit (e.g., \citealt{2010A&A...524A..25T}), we took spectra with CORALIE through a transit of WASP-111b on the night beginning 2013 Aug 29 (Figure~\ref{fig:w111-rm}). Over the sequence the airmass ranged over 1.69--1.01--1.09. 
The star was observed simultaneously with TRAPPIST (Figure \ref{fig:w111-rv-phot}).

A summary of our observations is presented in Table~\ref{tab:obs} and the follow-up photometry and RVs are provided in accompanying online tables; a guide to their content and format is given in Tables~\ref{tab:phot} and \ref{tab:rv}.

We analysed the WASP lightcurves of each star to determine whether they show 
periodic modulation due to the combination of magnetic activity and stellar 
rotation \citep{2011PASP..123..547M}. 
We found nothing above 1\,mmag for WASP-87, WASP-108 and WASP-109; the upper limit was 4\,mmag for WASP-110 and 2\,mmag for WASP-112. 
For WASP-111 we find nothing above 1\,mmag for the 2006 season; the upper limit was 10\,mmag for the 2007 season. 

During the aperture photometry of the EulerCam images of WASP-110 the aperture encompassed a fainter star. From 46 out-of-transit images with an average stellar FWHM of 1\farcs3, we measured a brightness ratio of $14.09 \pm 0.15$ and a separation of $4\farcs589 \pm 0\farcs006$ between WASP-110 and the nearby star. We corrected the EulerCam lightcurve for this dilution. 

\begin{table} 
\caption{Observations} 
\label{tab:obs}
\begin{tabular}{lcc}
\hline
Facility / filter	& Date			& N$_{\rm obs}$	\\
\hline
{\bf WASP-87:}	\\
WASP-South / broad	& 2011 Jan--Jul	& 8943	\\
Euler/CORALIE / spectros.	& 2012 May--Jul	& 12 RVs	\\
TRAPPIST / $z'$		& 2012 Jun 03	& 631	\\
TRAPPIST / $z'$		& 2012 Jun 08	& 857	\\
TRAPPIST / $z'$		& 2012 Jul 10	& 609	\\
EulerCam / Gunn $r$	& 2012 Jul 10	& 198	\\
EulerCam / Gunn $r$	& 2013 Feb 17	& 364	\\
Danish / R			& 2013 Apr 17	& 407	\\
TRAPPIST / $z'$		& 2013 Apr 27	& 895	\\
TRAPPIST / $z'$		& 2013 May 24	& 1051	\\
{\bf WASP-108:}	\\
WASP-South / broad	& 2011 Jan--2012 Jun	& 26769	\\
Euler/CORALIE / spectros.	& 2012 Jul--2013 Aug	& 23 RVs	\\
TRAPPIST / $V$		& 2013 Mar 05	& 729	\\
TRAPPIST / $Iz'$		& 2013 Mar 13	& 977	\\
EulerCam / Gunn $r$	& 2013 Mar 21	& 83	\\
TRAPPIST / $z'$		& 2013 Apr 06	& 1022	\\
Danish / R			& 2013 Apr 22	& 169	\\
EulerCam / Gunn $r$	& 2013 Jun 04	& 239	\\
TRAPPIST / $z'$		& 2014 May 15	& 1013	\\
{\bf WASP-109:}	\\
WASP-South / broad	& 2008 Jun--2012 Jun	& 23759	\\
Euler/CORALIE / spectros.	& 2012 May--2013 Aug	& 37 RVs	\\
EulerCam / Gunn $r$	& 2012 Jun 26	& 182	\\
TRAPPIST / $Iz'$		& 2013 Mar 05	& 542	\\
TRAPPIST / $Iz'$		& 2013 Apr 04	& 953	\\
EulerCam / $I$			& 2013 Apr 24	& 253	\\
Danish / R			& 2013 Apr 24	& 166	\\
TRAPPIST / $Iz'$		& 2013 May 04	& 857	\\
{\bf WASP-110:}	\\
WASP-South / broad	& 2006 May--2012 Jun	& 22698	\\
Euler/CORALIE / spectros.	& 2012 Jul--2013 Sep	& 14 RVs	\\
EulerCam / $I$			& 2013 Aug 16	& 121	\\
{\bf WASP-111:}	\\
WASP-South / broad	& 2006 May--2008 Oct	& 9386	\\
Euler/CORALIE / spectros.	& 2012 Jul--2012 Oct	& 28 RVs	\\
Euler/CORALIE / spectros.	& 2013 Jul--2013 Sep	& 12 RVs	\\
Euler/CORALIE / spectros.	& 2013 Aug 29			& 16 RVs	\\
TRAPPIST / $z'$		& 2012 Jul 25	& 684	\\
TRAPPIST / $B$		& 2012 Sep 21	& 562	\\
TRAPPIST / $B$		& 2012 Oct 28	& 535	\\
TRAPPIST / $z'$		& 2013 Aug 29	& 669	\\
EulerCam / Gunn $r$	& 2013 Nov 04	& 179	\\
{\bf WASP-112:}	\\
WASP-South / broad	& 2006 May--2011 Nov	& 14073	\\
Euler/CORALIE / spectros.	& 2012 Jul--2013 Sep	& 19 RVs	\\
TRAPPIST / blue blocking$^\dagger$	& 2012 Nov 27		& 556	\\
TRAPPIST / blue blocking$^\dagger$	& 2013 May 31		& 434	\\
TRAPPIST / blue blocking$^\dagger$	& 2013 Jun 03		& 609	\\
EulerCam / NGTS$^\ddagger$		& 2013 Aug 12	& 132	\\
EulerCam / NGTS$^\ddagger$		& 2013 Aug 15	& 223	\\
TRAPPIST / blue blocking$^\dagger$	& 2013 Aug 21		& 303	\\
TRAPPIST / blue blocking$^\dagger$	& 2013 Nov 02		& 426	\\
\hline
\end{tabular}
\newline $^\dagger$ \url{www.astrodon.com/products/filters/exoplanet}
\newline $^\ddagger$ \citet{2012SPIE.8444E..0EC}
\end{table}

\begin{table*} 
\caption{TRAPPIST, EulerCam and Danish photometry of the six stars} 
\label{tab:phot} 
\begin{tabular}{llllrrr} 
\hline 
\leftcell{Set} & \leftcell{Star} & \leftcell{Instrument} & Filter & \leftcell{BJD(UTC)} & \leftcell{Mag} & \leftcell{$\sigma_{\rm Mag}$}\\ 
    &      &      & & \leftcell{$-$2450000} &        &                 \\
& & & & \leftcell{(day)} & & \\
\hline
1	& WASP-87  & TRAPPIST & $z'$ & 6082.44789 & 0.009249 & 0.002876 \\
1	& WASP-87  & TRAPPIST & $z'$ & 6082.44812 & 0.008878 & 0.002866 \\
\ldots \\
\ldots \\
43	& WASP-112 & TRAPPIST & blue-blocking & 6599.64687 &    0.005878 & 0.005179 \\
43	& WASP-112 & TRAPPIST & blue-blocking & 6599.64716 & $-$0.005336 & 0.005165 \\
\hline
\end{tabular}
\begin{flushleft}
The magnitude values are differential and normalised to the out-of-transit levels. 
The contamination of the WASP-110 photometry was accounted for. 
\newline The uncertainties are the formal errors (i.e. they have not been rescaled).
\newline This table is available in its entirety via the CDS. 
\end{flushleft}
\end{table*}

\begin{table}
\caption{Radial velocity measurements} 
\label{tab:rv} 
\begin{tabular*}{0.48\textwidth}{lrrrrr} 
\hline 
\leftcell{Star} & \leftcell{BJD(UTC)} & \leftcell{RV} & \leftcell{$\sigma$$_{\rm RV}$} & \leftcell{BS}\\ 
     & \leftcell{$-$2450000}&    &                     &   \\
     & \leftcell{(day)}     & \leftcell{(km s$^{-1}$)} & \leftcell{(km s$^{-1}$)} & \leftcell{(km s$^{-1}$)}\\ 
\hline
WASP-87  & 6068.56103 & $-$13.8988 & 0.0234 &    0.0472 \\
WASP-87  & 6069.51421 & $-$14.4770 & 0.0366 & $-$0.0319 \\
\ldots\\
WASP-112 & 6537.81718 & $-$19.4170 & 0.0904 & $-$0.0498 \\
WASP-112 & 6543.54834 & $-$19.6296 & 0.0546 &    0.0312 \\
\hline
\end{tabular*}
The uncertainties are the formal errors (i.e. with no added jitter) and the WASP-111 RVs from 2012 have not been pre-whitened.\\  
The uncertainties on bisector span (BS) is 2\,$\sigma_{\rm RV}$.\\
This table is available in its entirety via the CDS. 
\end{table}

\clearpage

\clearpage

\clearpage

\section{Stellar parameters from spectra}
\label{sec:stellar}
The individual CORALIE spectra of the host stars were co-added to produce a single spectrum per star. 
We performed the spectral analysis using the methods given in \citet{2013MNRAS.428.3164D}. The excitation balance of the Fe~{\sc i} lines was used to determine the effective temperature (\teff). The surface gravity (\logg) was determined from the ionisation balance of Fe~{\sc i} and Fe~{\sc ii}. The Ca~{\sc i} line at 6439{\AA} and the Na~{\sc i} D lines were also used as \logg\ diagnostics. The elemental abundances were determined from equivalent width measurements of several unblended lines. The quoted error estimates include that given by the uncertainties in \teff\ and \logg, as well as the scatter due to measurement and atomic data uncertainties.
The projected stellar rotation velocity (\vsini) was determined by fitting the profiles of several unblended Fe~{\sc i} lines. Values of macroturbulent velocity were assumed from the asteorseismic-based calibration of \citet{2014MNRAS.444.3592D}.
The parameters obtained from the analysis are listed in Table~\ref{tab:params}.

\section{System parameters from combined analyses}
\label{sec:mcmc}
We determined the parameters of each system from a simultaneous fit to the lightcurve and radial-velocity data. 
The fit was performed using the current version of the 
Markov-chain Monte Carlo (MCMC) code described by \citet{2007MNRAS.380.1230C} 
and \citet{2014arXiv1402.1482A}.

The transit lightcurves were modelled using the formulation of 
\citet{2002ApJ...580L.171M} and limb-darkening was accounted for using the four-parameter non-linear law of \citet{2000A&A...363.1081C, 2004A&A...428.1001C} (see Table~\ref{tab:ld} for the interpolated coefficients). 
The RM effect was modelled using the formulation of \citet{2011ApJ...742...69H}.
Stellar density is determined by the transit lightcurves, but we require a constraint on stellar mass for a full characterisation of the system. For this we used the empirical mass calibration of \citet[see also the references therein]{2011MNRAS.417.2166S}.

We used the $F$-test approach of \citet{1971AJ.....76..544L} to calculate the probability that the improvement in the fit that results from fitting an eccentric orbit could have arisen by chance if the underlying orbit were circular. The values range from 0.33 (for WASP-112) to 0.93 (for WASP-87); for no system is there compelling evidence of a non-circular orbit. 
We thus adopt circular orbits, which \citet{2012MNRAS.422.1988A} argue is the prudent choice for short-period,$\sim$Jupiter-mass planets in the absence of evidence to the contrary.

The fitted parameters were $T_{\rm 0}$, $P$, (\rplanet/\rstar)$^2$, $T_{\rm 14}$, $b$, $K_{\rm 1}$, $\gamma$, \svsicos, \svsisin, \teff\ and \feh, where $T_{\rm 0}$ is the epoch of mid-transit , $P$, is the orbital period, (\rplanet/\rstar)$^2$ is the planet-to-star area ratio, $T_{\rm 14}$ is the total transit duration, $b$ is the impact parameter of the planet's path across the stellar disc, $K_{\rm 1}$ is the reflex velocity semi-amplitude, $\gamma$ is the systemic velocity, \vsini\ is the sky-projected stellar rotation velocity, $\lambda$ is the sky-projected obliquity, \teff\ is the stellar effective temperature and \feh\ is the stellar metallicity. We placed priors on both \teff\ and \feh\ using the spectroscopically-derived values.

To give proper weighting to each photometric data set, the uncertainties were 
scaled so as to obtain a photometric reduced-\chisq\ of unity. 
To allow for flux offsets, we partitioned those TRAPPIST lightcurves involving meridian flips and the EulerCam $r$-band transit of WASP-109b, which was interrupted by a power cut. 

\begin{figure}
\includegraphics[width=90mm]{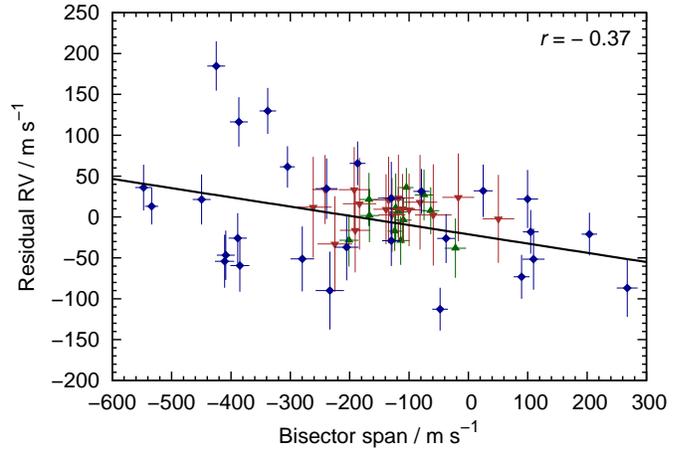}
\caption{The correlation between residual RV and bisector span for WASP-111. 
The symbol shapes and colours are as in Figure~\ref{fig:w111-rv-phot}.
The fit and subsequent correction was made to the 2012 data alone; the data from 2013 are shown for reference.
\label{fig:w111-res-bis}} 
\end{figure}

To obtain a spectroscopic reduced-$\chi^2$ of unity we added a `jitter' term of 52\,\ms\ in quadrature to the formal RV errors of WASP-109. No such addition was required in the cases of WASP-87, -108, -110 or -112. 
The RVs of WASP-111 were partitioned into three sets, each with its own fitted $\gamma$: the 2012 data, the 2013 data and the RM sequence. 
We observed greater scatter in the star's RVs from 2012, perhaps indicating a period of greater activity. 
We pre-whitened those data using the correlation apparent between the residual RV and bisector span (Figure~\ref{fig:w111-res-bis}). This resulted in a drop in RMS for the 2012 data from 68.4\,\ms\ to 63.4\,\ms.
We found an offset between the corrected 2012 data and the 2013 data of $-42.56 \pm 0.90$\,\ms\ and an offset of $22.0 \pm 8.8$\,\ms\ between the 2013 RM sequence and the 2013 data.
The jitter terms for the WASP-111 data were 55\,\ms\ for 2012, 15\,\ms\ for 2013 and zero for the RM sequence. 
During an initial MCMC we found \vsini, from the fit to the RM effect, to range over 0--24\,\kms\ (with a best-fitting value of $9.5^{+3.4}_{-4.0}$\,\kms), which is inconsistent with the spectroscopic value of $11.2 \pm 0.8$\,\kms. We thus used the spectroscopic value to place a prior on \vsini, though we note that this barely affected the derived spin-orbit angle: we obtained $\lambda = -5 \pm 16^{\circ}$ with a prior and $\lambda = -5 \pm 19^{\circ}$ without a prior.

\begin{table*}
\centering
\caption{Limb-darkening coefficients} 
\label{tab:ld} 
\begin{tabular}{llllllll}
\hline
\leftcell{Star}		& \leftcell{Instrument}		& \leftcell{Observation bands}		& \leftcell{Claret band}			& \leftcell{$a_1$}		& \leftcell{$a_2$}		& \leftcell{$a_3$}		& \leftcell{$a_4$}	\\
\hline
WASP-87		& WASP / EulerCam / Danish	& Broad (400--700 nm) / Gunn $r$ / Cousins R & Cousins $R$ & 0.358 & 0.740 & $-$0.600 & 0.165 \\
WASP-87		& TRAPPIST	& Sloan $z'$ & Sloan $z'$ & 0.447 & 0.282 & $-$0.190 & 0.012 \\
WASP-108	& WASP / EulerCam / Danish	& Broad (400--700 nm) / Gunn $r$ / Cousins R & Cousins $R$ & 0.593 & $-$0.049 & 0.464 & $-$0.274 \\
WASP-108	& TRAPPIST	& Cousins $I$ + Sloan $z'$ / Sloan $z'$ & Sloan $z'$ &  0.685 & $-$0.401 & 0.668 & $-$0.331 \\
WASP-108	& TRAPPIST	& Johnson $V$ & Johnson $V$ & 0.514 & 0.145 & 0.383 & $-$0.248 \\
WASP-109	& WASP / EulerCam / Danish	& Broad (400--700 nm) / Gunn $r$ / Cousins R & Cousins $R$ & 0.499 & 0.396 & $-$0.226 & 0.025 \\
WASP-109	& TRAPPIST	& Cousins $I$ + Sloan $z'$ & Sloan $z'$ & 0.589 & $-$0.027 & 0.120 & $-$0.010 \\
WASP-109	& EulerCam	& Cousins $I$ & Cousins $I$ & 0.569 & 0.093 & 0.025 & $-$0.066 \\
WASP-110	& WASP & Broad (400--700 nm) & Cousins $R$ & 0.657 & $-$0.456 & 1.100 & $-$0.529 \\
WASP-110	& EulerCam	& Cousins $I$ & Cousins $I$ & 0.737 & $-$0.666 & 1.156 & $-$0.529 \\
WASP-111	& WASP / EulerCam & Broad (400--700 nm) / Gunn $r$ & Cousins $R$ & 0.488 & 0.454 & $-$0.310 & 0.064 \\
WASP-111	& TRAPPIST	& Sloan $z'$ & Sloan $z'$ & 0.582 & 0.017 & 0.055 & $-$0.071 \\
WASP-111	& TRAPPIST	& Johnson $B$ & Johnson $B$ & 0.344 & 0.676 & $-$0.178 & $-$0.003 \\
WASP-112	& WASP / EulerCam / TRAPPIST & Broad (400--700 nm) / NGTS / blue block. & Cousins $R$ & 0.449 & 0.256 & 0.196 & $-$0.188 \\
\hline
\end{tabular}
\end{table*}


\section{Comparison to evolutionary models}
\label{sec:evol}
We used a Markov chain Monte Carlo method to calculate the posterior
distribution for the  mass and age  of each host star based on a comparison
of the observed values of \rhostar, \Teff\ and [Fe/H] to a grid of stellar
models.  The method is described in detail in \citet{bagemass}. The
stellar models were calculated using the {\sc garstec} stellar evolution code
\citep{2008Ap&SS.316...99W} and the methods used to calculate the stellar
model grid are described in \citet{2013MNRAS.429.3645S}.

 To estimate the likelihood of observing the data \mbox{$\bmath{d}=
\left(\Teff, \rmn{[Fe/H]},\rhostar\right)$} for a given model \mbox{$\bmath{m}
= \left(M_{\star},\taustar,{\rm [Fe/H]}\right)$} (where \taustar\ is the age
of the star) we use \mbox{${\cal L}(\bmath{d}|\bmath{m}) = \exp(-\chi^2/2)$},
where
\[
\chi^2 =  \frac{\left(T_{\rmn{eff}} - T_{\rmn{eff,obs}}\right)^2}{\sigma_T^2} 
+ \frac{\left(\rmn{[Fe/H]}_{\rmn{s}} - \rmn{[Fe/H]}_{\rmn{s,obs}}\right)^2}
        {\sigma_{\rmn{[Fe/H]}}^2}\\
  + \frac{\left(\rho_{\star} - \rho_{\star,\rmn{obs}}\right)^2}{\sigma_{\rho}^2}.
\]
 In this expression for $\chi^2$ observed quantities are denoted by the
subscript `obs', their standard errors are $\sigma_T$, etc., and other
quantities are derived from the model grid using spline interpolation. We use
the Metropolis-Hastings algorithm to produce a Markov chain  of 50,000 steps
with the Bayesian probability distribution \mbox{$p(\bmath{m}|\bmath{d})$}
\citep{2004PhRvD..69j3501T}. We assume uniform priors for the mass, age and
composition of the star over the range of valid grid values ($M_{\star} =
0.6$\,\msol\ to 2.0\,\msol, $\rmn{[Fe/H]} = -0.75$ to  0.55 and $\taustar
= 0.1$\,Gyr  to either 17.5\,Gyr or the age at which the stellar radius first
exceeds 3\rsol). The results of this Bayesian analysis are given in Table~\ref{AgeMassTable}; 
there is good agreement with the stellar masses derived in the MCMC analyses (compare with Table~\ref{tab:params}). 

\begin{figure}
\mbox{\includegraphics[width=0.45\textwidth]{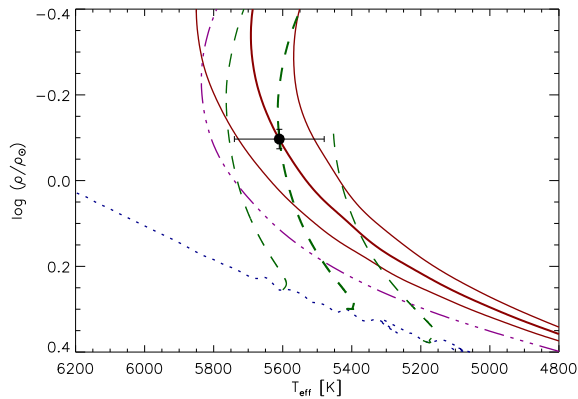}}
\caption{Mean stellar density versus effective temperature for WASP-112
compared to stellar evolution models assuming $\alpha_{\rm MLT}=1.22$ and the
best-fitting value of [Fe/H]. Isochrones (solid lines, red) are shown for the
best-fitting age and also for age $\pm 1$-$\sigma$. Similarly, evolution tracks
(dashed lines, green) are shown for the best-fitting mass and $\pm 1$-$\sigma $. The zero-age main sequence is also shown (dotted line, cyan). For comparision, an isochrone is shown assuming the standard value of $\alpha_{\rm MLT}=1.78$ and
an age of 14.8\,Gyr (dot-dash line, magenta). 
\label{fig:trho112}}
\end{figure}

For WASP-112, using our standard grid of stellar models we found that the
probability that the age of this star is less than the age of the Galactic
disc \citep[10\,Gyr,][]{2014A&A...566A..81C} is $p(\taustar < 10\,\rmn{Gyr}) =
0.04$. This makes it likely that this star is affected by the ``radius
anomaly'': some late-type stars appear to be significantly larger than predicted by standard stellar models \citep{1973A&A....26..437H, 1997AJ....114.1195P, 2007ApJ...660..732L, 2013ApJ...776...87S}. 
It has been
proposed that this is due to the magnetic field reducing the efficiency of
energy transport by convection, a phenomenon that can be approximated by
reducing the mixing length parameter used in the model
\citep{2013ApJ...779..183F, 2007A&A...472L..17C}. The mixing length parameter
used to calculate our model grid is $\alpha_{\rm MLT}=1.78$. With this value
of $\alpha_{\rm MLT}$ {\sc garstec}  reproduces the observed properties of the
present day Sun.  There is currently no practical way to select the correct
value of $\alpha_{\rm MLT}$ for a magnetically active star other than to find
the range of this parameter that gives plausible results. Accordingly, we
calculated a Markov chain for the observed parameters of WASP-112 using
stellar models with $\alpha_{\rm MLT}=1.22$, for which value we find
$p(\taustar < 10\,\rmn{Gyr}) = 0.43$. The fit to the observed density and
effective temperature of WASP-112 is shown in Fig.~\ref{fig:trho112}.

\begin{table}
 \caption{Bayesian mass and age estimates for the host stars  Columns 2 and 3 give the maximum-likelihood estimates of the age and
mass, respectively. The chi-squared value for the the best fit is given
in column 4.  Columns 5 and 6 give the mean and standard deviation of the 
posterior age and mass distribution, respectively.
\label{AgeMassTable}}
 \begin{tabular}{@{}lrrrrr}
\hline
  \multicolumn{1}{@{}l}{Star} &
  \multicolumn{1}{l}{$\tau_{\rm b}$} &
  \multicolumn{1}{l}{$M_{\rm b}$} &
  \multicolumn{1}{l}{$\chi^2$}&
  \multicolumn{1}{l}{$\langle \taustar \rangle$ }  &
  \multicolumn{1}{l}{$\langle \Mstar \rangle$ }\\
  \multicolumn{1}{l}{} &
  \multicolumn{1}{l}{[Gyr]} &
  \multicolumn{1}{l}{[\msol]} &
  \multicolumn{1}{l}{}&
  \multicolumn{1}{l}{[Gyr]}  &
  \multicolumn{1}{l}{[\msol]}\\
\hline
 \noalign{\smallskip}
WASP-87      & 3.4 & 1.20 & 0.05 &$  3.8 \pm 0.8  $&$ 1.20 \pm 0.08  $\\
WASP-108     & 4.3 & 1.11 & 0.21 &$  4.6 \pm 1.9  $&$ 1.10 \pm 0.07  $\\
WASP-109     & 2.4 & 1.20 & 0.27 &$  2.6 \pm 0.9  $&$ 1.19 \pm 0.06  $\\
WASP-110     & 8.4 & 0.87 & 0.29 &$  8.6 \pm 3.5  $&$ 0.87  \pm 0.05  $\\
WASP-111     & 2.6 & 1.40 & 0.18 &$  2.6 \pm 0.6  $&$ 1.41  \pm 0.07  $\\
WASP-112\parbox{0pt}{$^a$} & 12.5 & 0.80 & 0.08 &$ 10.6 \pm 3.0  $&$ 0.83 \pm 0.05 $\\
 \noalign{\smallskip}
\hline
\multicolumn{6}{@{}l}{$^a$ Assuming $\alpha_{\rm MLT}=1.22$}
 \end{tabular}   
 \end{table}     

\section{The planetary systems}

\subsection{The WASP-87 system}
WASP-87b is a 2.2-\mjup, 1.39-\rjup\ planet in a 1.68-day orbit around a $V = 10.6$, metal-poor F5 star. 
With \feh\ = $-0.41$, WASP-87 is one of the most metal-poor stars found to host a giant planet. 
From \vsini\ and \rstar\ we calculated a stellar rotation period of \prot\ $< 8.6 \pm 0.7$\,d. Together with the orbital period, \porb, of 1.68\,d, this places WASP-87 in the sparsely-populated region of \prot-\porb\ space identified by \citet{2013ApJ...775L..11M}. 
The radius we derived for WASP-87b is smaller by 0.27\,\rjup\ than predicted by the empirical relation of \citet{2012A&A...540A..99E} based on the planet's \teql, \mplanet\ and \feh. The average difference between the predicted and observed radii for the calibration sample of \citet{2012A&A...540A..99E} was 0.11\,\rjup. 

\begin{figure}
\includegraphics[width=90mm]{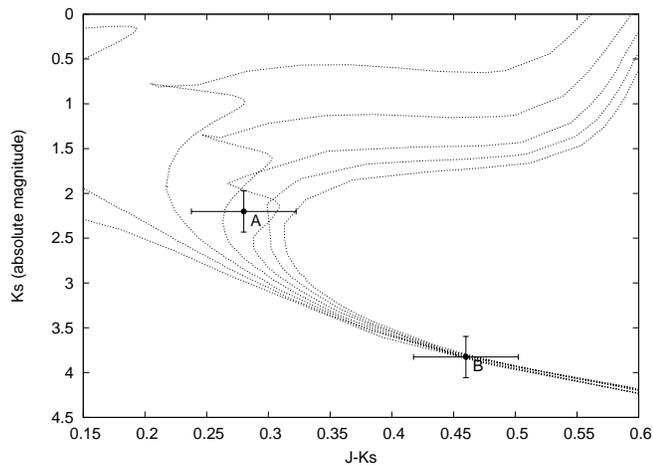}
\caption{Colour-magnitude diagram for WASP-87A+B. 
The isochrones range from 0 to 6 Gyr in 1-Gyr increments \citep{2008A&A...482..883M}.
\label{fig:w87-cmd}} 
\end{figure}

There is another star, 2MASS\,12211848$-$5250332, located 8\farcs2 to the south-east of WASP-87A with $V_{\rm mag} = 12.8$ and $K_{\rm mag} = 11.2$. From its spectral energy distribution we found \teff\ = $5700 \pm 150$ K. The nearby star's proper motion in UCAC4 ($\mu_{\rm RA} = 1.3 \pm 1.7$ mas/yr, $\mu_{\rm Dec} = -2.9 \pm 2.0$ mas/yr) and that of WASP-87A ($\mu_{\rm RA} = -3.4 \pm 0.8$ mas/yr, $\mu_{\rm Dec} = 13.9 \pm 2.1$ mas/yr) are similar. From three CORALIE spectra of the nearby star we measured its radial velocity to be $-13.348 \pm 0.013$\,\kms, which is close to the systemic velocity of WASP-87A ($-14.1845 \pm 0.0079$\,\kms). We interpret this as suggesting that WASP-87A and the nearby star, which we tentatively name WASP-87B, comprise a bound system. We constructed a colour-magnitude diagram using 2MASS magnitudes that indicates a system age of 3--4\,Gyr (Figure~\ref{fig:w87-cmd}). This is in good agreement with the age of $3.8 \pm 0.6$\,Gyr that we derived from an evolutionary analysis.

\subsection{The WASP-108 system}
WASP-108b is a 0.89-\mjup, 1.28-\rjup\ planet in a 2.68-day orbit around a $V = 11.2$, \feh\ = $+0.05$, F9 star. 
From an evolutionary analysis we find the stellar age to be $4.6 \pm 1.9$\,Gyr. 

\subsection{The WASP-109 system}
WASP-109b is a 0.9-\mjup, 1.44-\rjup\ planet in a 3.32-day orbit around a $V = 11.4$, \feh\ = $-0.22$, F4 star. 
Our evolutionary analysis suggests an age of $2.6 \pm 0.9$\,Gyr. 
From \vsini\ and \rstar\ we calculated a stellar rotation period of \prot\ $< 4.4 \pm 0.3$\,d. Together with the orbital period, \porb, of 3.32\,d, this places WASP-109 in the sparsely-populated region of \prot-\porb\ space identified by \citet{2013ApJ...775L..11M}.

\subsection{The WASP-110 system}
WASP-110b is a 0.51-\mjup, 1.24-\rjup\ planet in a 3.78-day orbit around a $V = 12.3$, \feh\ = $-0.06$, G9 star. 
Our evolutionary analysis suggests a stellar age of $8.6 \pm 3.5$\,Gyr. 
Our radius for WASP-110b (1.38\,\rjup) is smaller by 0.22\,\rjup\ than predicted by the empirical relation of \citet{2012A&A...540A..99E}
There is a star 14 times fainter than WASP-110 separated by 4\farcs6. 
Further observations are required to ascertain whether they comprise a visual binary or are merely a line-of-sight coincidence. 

\subsection{The WASP-111 system}
WASP-111b is a 1.83-\mjup, 1.44-\rjup\ planet in a 2.31-day orbit around a $V = 10.3$, \feh\ = $+0.08$, F5 star. 
Our evolutionary analysis suggests an age of $2.6 \pm 0.6$\,Gyr.
With \prot\ $< 8.4 \pm 0.7$\,d and \porb\ = 2.31\,d, WASP-111 also occupies the  sparsely-populated region of \prot-\porb\ space identified by \citet{2013ApJ...775L..11M}. 
Similar to WASP-87b and WASP-110b, we derived a radius for WASP-111b that is 0.23\,\rjup\ smaller than predicted by the empiricial relation of \citet{2012A&A...540A..99E}. The masses and irradiation levels of both WASP-87b and WASP-11b are similar, but their metallicities differ. 

From a fit to the RM effect (Figure~\ref{fig:w111-rm}), we find WASP-111b to be in a prograde orbit, 
with a sky-projected stellar obliquity $\lambda$ of $-5 \pm 16^\circ$. 
If the stellar spin axis is near-aligned with the sky plane then WASP-111b is in a near-aligned orbit, 
i.e. the true obliquity $\Psi \sim \lambda$. 
As \teff\ = $6400 \pm 150$\,K, WASP-111 is expected to have little convective envelope and so to weakly tidally interact with 
WASP-111b \citep{2010ApJ...718L.145W}.
This would suggest either that the planet migrated via alignment-preserving tidal interaction with a protoplanetary disc or via scattering processes that, by chance, left the planet in a near-aligned orbit (e.g. \citealt{2014MNRAS.445.1114A}).

\begin{figure*}
\includegraphics[width=168mm]{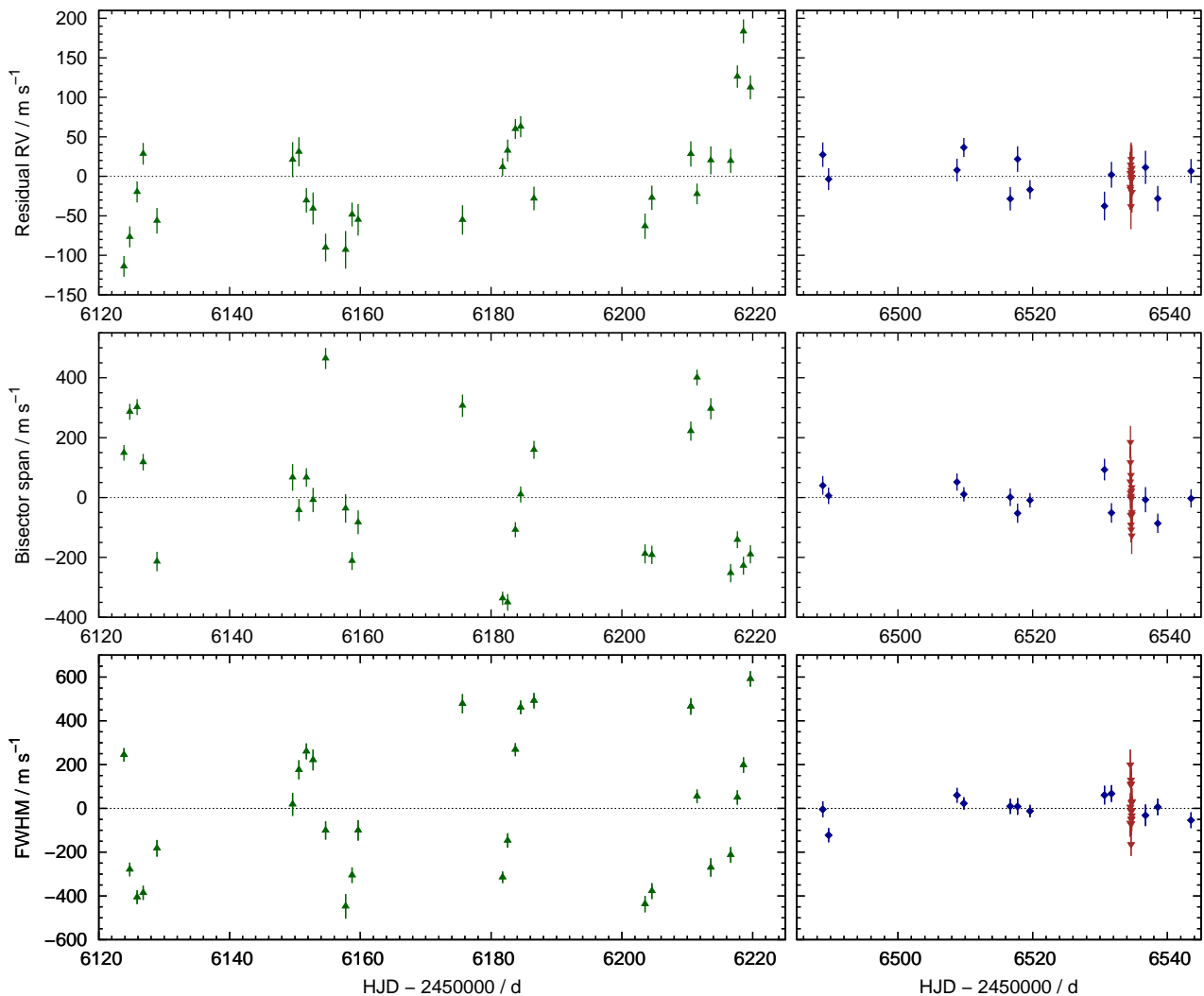}
\caption{The variable activity of WASP-111. The panels to the left show the data from 2012 Jul 14--Oct 19 and the panels to the right show the data from 2013 Jul 14--Sep 8, including the transit observation of Aug 29. 
The symbol shapes and colours are as in Figure~\ref{fig:w111-rv-phot}.
The weighted mean was subtracted from each dataset for the bisector span and FWHM plots. 
{\it Top panels:} Residuals about a fit to the RVs of a circular Keplerian orbit. The RVs were partitioned into three sets: 2012 (not pre-whitened), 2013 and the RM effect. 
{\it Middle panels:} Bisector span of the CCF. 
{\it Bottom panels:} FWHM of the CCF. 
\label{fig:w111-activity}} 
\end{figure*}

It appears that WASP-111 was active when we observed it with CORALIE in July to October of 2012, then quiescent during July to September of the following year. 
The scatter in the bisector span and the FWHM of the CCF, and in the RV residuals was much larger in 2012 than in 2013 (Figure~\ref{fig:w111-activity}).
This indicates that WASP-111 (\teff\ $\sim$ 6400 K) undergoes activity cycles like the Sun, and may have a shorter cycle period like $\tau$ Boo (\teff\ $\sim$ 6300\,K; \citealt{2004A&A...418..989N}). 
In 2012, the peak to peak variation in residual RV, CCF bisector span and CCF FWHM was, respectively, 297\,\ms, 815\,\ms\ and 1039\,\ms. 
The corresponding values during the RM sequence of 2013 Aug 29 were factors of a few smaller: 60\,\ms, 312\,\ms\ and 364\,\ms. 
Perhaps this information, together with the size of the planet, could be used to determine the effective size and intensity of the active regions (e.g. \citealt{2014arXiv1409.3594D}).
The variable nature of the stellar activity of WASP-111, the star's brightness and the planet's close orbit make it an ideal candidate for studying the effects of variable activity on a hot-Jupiter atmosphere. 

\subsection{The WASP-112 system}
WASP-112b is a 0.9-\mjup, 1.19-\rjup\ planet in a 3.04-day orbit around a $V = 13.3$, metal-poor star (\feh\ = $-0.64$). The \teff\ of 5610\,K implies a spectral type of G6, though our derived stellar mass of 0.81\,\msol\ is suggestive of K0V. 
As a check of our spectroscopic \teff, we used the infrared flux method (IRFM) of \citet{1977MNRAS.180..177B} and obtained \teff\ = $5760 \pm 130$ K, which is even more discrepant with our derived stellar mass. 

The low density of WASP-112 cannot be matched by standard stellar models for
any reasonable value of age (Section~\ref{sec:evol}). 
This could be due to unreliable spectral parameters as, due to the faintness of the star, the S/N of the coadded CORALIE spectrum is only $\sim$50:1.
We found that the MCMC method of \citet{bagemass} gave a good fit for \feh\ = $0.2 \pm 0.2$ (cf. the spectroscopic value of $-0.64 \pm 0.15$) when using the IRFM \teff, \rhostar\ from Table~\ref{tab:params}, and with an upper limit on the age of $10 \pm 1$ Gyr. 
Alternatively, we can match the properties of
this star if we assume a much lower value for the mixing length parameter in
the models than the value $\amlt=1.78$ that comes from calibrating the models
using the present-day properties of the Sun. There is some evidence that low
values of \amlt\ are appropriate for models of K-dwarfs like WASP-112 if they
are magnetically active. Magnetic activity is a consequence of rapid rotation
in low mass stars and can be detected through chromospheric emission lines,
e.g., Ca\,{\sc ii} H- and K-line emission, or modulation of the light curve due to
star spots. However, WASP-112 does not show any strong signal of star spot
activity in its light curve, and the projected rotation
velocity of this star is very low. 
We can exclude strong Ca\,{\sc ii} H+K emission, but the CORALIE spectra are too low S/N to be sensitive to weak emission. 
It may be that WASP-112 is a rapidly rotating star, but that the  rotation axis 
is oriented towards the Earth so that the
projected rotational velocity is low and the modulation in brightness due to
the changing visibility of star spots with rotation is negligible. If this is
the case, then the orbit of WASP-112b must be misaligned with the rotation
axis of the star, i.e, WASP-112b is on a polar orbit around WASP-112. This
hypothesis can be directly tested using observations of the
RM effect.

WASP-112 has the has lowest metallicity of any star known to be orbited by a Jupiter-like planet; with \feh\ = $-$0.60, WASP-98 is the next most metal-poor giant-planet host. If stellar metallicity is representative of protoplanetary-disc metallicity, the discs of these two stars would have contained relatively little solid material.
Further, both stars are low mass (WASP-98 has $0.69 \pm 0.06$\,\msol\ and WASP-112 has $0.81 \pm 0.07$\,\msol) so their protoplanetary discs are expected to have been relatively low mass. 
Both the low metallicity and the low mass of these two stars may pose problems for the model of formation of giant planets via core accretion.

\section*{Acknowledgements}
WASP-South is hosted by the South African Astronomical Observatory; we are grateful for their ongoing support and assistance. 
Funding for WASP comes from consortium universities and from the UK's Science and Technology Facilities Council. 
The Euler-Swiss telescope is supported by the Swiss National Science Foundation.
TRAPPIST is funded by the Belgian Fund for Scientific Research (Fond National de la Recherche Scientifique, FNRS) under the grant FRFC 2.5.594.09.F, with the participation of the Swiss National Science Fundation (SNF). 
L. Delrez is a FNRS/FRIA Doctoral Fellow. 
M. Gillon and E. Jehin are FNRS Research Associates.
P. Rojo is supported by Fondecyt grant \#1120299.
A. Serenelli acknowledges support by the MICINN grants AYA2011-24704 and ESP2013-41268-R. 
A. H. M. J. Triaud is a Swiss National Science Foundation fellow under grant number P300P2-147773.



\begin{thebibliography}{52}
\expandafter\ifx\csname natexlab\endcsname\relax\def\natexlab#1{#1}\fi

\bibitem[{{Anderson} {et~al}\mbox{.}(2014){Anderson}, {Collier Cameron},
  {Delrez}, {Doyle}, {Faedi}, {Fumel}, {Gillon}, {G{\'o}mez Maqueo Chew},
  {Hellier}, {Jehin}, {Lendl}, {Maxted}, {Pepe}, {Pollacco}, {Queloz},
  {S{\'e}gransan}, {Skillen}, {Smalley}, {Smith}, {Southworth}, {Triaud},
  {Turner}, {Udry}, \& {West}}]{2014MNRAS.445.1114A}
{Anderson} D.~R. {et~al.}, 2014, \mnras, 445, 1114

\bibitem[{{Anderson} {et~al}\mbox{.}(2012){Anderson}, {Collier Cameron},
  {Gillon}, {Hellier}, {Jehin}, {Lendl}, {Maxted}, {Queloz}, {Smalley},
  {Smith}, {Triaud}, {West}, {Pepe}, {Pollacco}, {S{\'e}gransan}, {Todd}, \&
  {Udry}}]{2012MNRAS.422.1988A}
---, 2012, \mnras, 422, 1988

\bibitem[{{Anderson} {et~al}\mbox{.}(2014){Anderson}, {Collier Cameron},
  {Hellier}, {Lendl}, {Lister}, {Maxted}, {Queloz}, {Smalley}, {Smith},
  {Triaud}, {Brown}, {Gillon}, {Neveu-VanMalle}, {Pepe}, {Pollacco},
  {Segransan}, {Udry}, {West}, \& {Wheatley}}]{2014arXiv1402.1482A}
---, 2014, \aap, submitted, arXiv:1402.1482

\bibitem[{{Asplund} {et~al}\mbox{.}(2009){Asplund}, {Grevesse}, {Sauval}, \&
  {Scott}}]{2009ARA&A..47..481A}
{Asplund} M., {Grevesse} N., {Sauval} A.~J., {Scott} P., 2009, \araa, 47, 481

\bibitem[{{Blackwell} \& {Shallis}(1977)}]{1977MNRAS.180..177B}
{Blackwell} D.~E., {Shallis} M.~J., 1977, \mnras, 180, 177

\bibitem[{{Borucki} {et~al}\mbox{.}(2010){Borucki}, {Koch}, {Basri}, {Batalha},
  {Brown}, {Caldwell}, {Caldwell}, {Christensen-Dalsgaard}, {Cochran},
  {DeVore}, {Dunham}, {Dupree}, {Gautier}, {Geary}, {Gilliland}, {Gould},
  {Howell}, {Jenkins}, {Kondo}, {Latham}, {Marcy}, {Meibom}, {Kjeldsen},
  {Lissauer}, {Monet}, {Morrison}, {Sasselov}, {Tarter}, {Boss}, {Brownlee},
  {Owen}, {Buzasi}, {Charbonneau}, {Doyle}, {Fortney}, {Ford}, {Holman},
  {Seager}, {Steffen}, {Welsh}, {Rowe}, {Anderson}, {Buchhave}, {Ciardi},
  {Walkowicz}, {Sherry}, {Horch}, {Isaacson}, {Everett}, {Fischer}, {Torres},
  {Johnson}, {Endl}, {MacQueen}, {Bryson}, {Dotson}, {Haas}, {Kolodziejczak},
  {Van Cleve}, {Chandrasekaran}, {Twicken}, {Quintana}, {Clarke}, {Allen},
  {Li}, {Wu}, {Tenenbaum}, {Verner}, {Bruhweiler}, {Barnes}, \&
  {Prsa}}]{2010Sci...327..977B}
{Borucki} W.~J. {et~al.}, 2010, Science, 327, 977

\bibitem[{{Chabrier}, {Gallardo} \& {Baraffe}(2007){Chabrier}, {Gallardo}, \&
  {Baraffe}}]{2007A&A...472L..17C}
{Chabrier} G., {Gallardo} J., {Baraffe} I., 2007, \aap, 472, L17

\bibitem[{{Chazelas} {et~al}\mbox{.}(2012){Chazelas}, {Pollacco}, {Queloz},
  {Rauer}, {Wheatley}, {West}, {Da Silva Bento}, {Burleigh}, {McCormac},
  {Eigm{\"u}ller}, {Erikson}, {Genolet}, {Goad}, {Jord{\'a}n}, {Neveu}, \&
  {Walker}}]{2012SPIE.8444E..0EC}
{Chazelas} B. {et~al.}, 2012, in Society of Photo-Optical Instrumentation
  Engineers (SPIE) Conference Series, Vol. 8444, Society of Photo-Optical
  Instrumentation Engineers (SPIE) Conference Series

\bibitem[{{Claret}(2000)}]{2000A&A...363.1081C}
{Claret} A., 2000, \aap, 363, 1081

\bibitem[{{Claret}(2004)}]{2004A&A...428.1001C}
---, 2004, \aap, 428, 1001

\bibitem[{{Cojocaru} {et~al}\mbox{.}(2014){Cojocaru}, {Torres}, {Isern}, \&
  {Garc{\'{\i}}a-Berro}}]{2014A&A...566A..81C}
{Cojocaru} R., {Torres} S., {Isern} J., {Garc{\'{\i}}a-Berro} E., 2014, \aap,
  566, A81

\bibitem[{{Collier Cameron} {et~al}\mbox{.}(2010){Collier Cameron}, {Guenther},
  {Smalley}, {McDonald}, {Hebb}, {Andersen}, {Augusteijn}, {Barros}, {Brown},
  {Cochran}, {Endl}, {Fossey}, {Hartmann}, {Maxted}, {Pollacco}, {Skillen},
  {Telting}, {Waldmann}, \& {West}}]{2010MNRAS.407..507C}
{Collier Cameron} A. {et~al.}, 2010, \mnras, 407, 507

\bibitem[{{Collier Cameron} {et~al}\mbox{.}(2006){Collier Cameron}, {Pollacco},
  {Street}, {Lister}, {West}, {Wilson}, {Pont}, {Christian}, {Clarkson},
  {Enoch}, {Evans}, {Fitzsimmons}, {Haswell}, {Hellier}, {Hodgkin}, {Horne},
  {Irwin}, {Kane}, {Keenan}, {Norton}, {Parley}, {Osborne}, {Ryans}, {Skillen},
  \& {Wheatley}}]{2006MNRAS.373..799C}
---, 2006, \mnras, 373, 799

\bibitem[{{Collier Cameron} {et~al}\mbox{.}(2007){Collier Cameron}, {Wilson},
  {West}, {Hebb}, {Wang}, {Aigrain}, {Bouchy}, {Christian}, {Clarkson},
  {Enoch}, {Esposito}, {Guenther}, {Haswell}, {H{\'e}brard}, {Hellier},
  {Horne}, {Irwin}, {Kane}, {Loeillet}, {Lister}, {Maxted}, {Mayor}, {Moutou},
  {Parley}, {Pollacco}, {Pont}, {Queloz}, {Ryans}, {Skillen}, {Street}, {Udry},
  \& {Wheatley}}]{2007MNRAS.380.1230C}
---, 2007, \mnras, 380, 1230

\bibitem[{{Doyle} {et~al}\mbox{.}(2014){Doyle}, {Davies}, {Smalley}, {Chaplin},
  \& {Elsworth}}]{2014MNRAS.444.3592D}
{Doyle} A.~P., {Davies} G.~R., {Smalley} B., {Chaplin} W.~J., {Elsworth} Y.,
  2014, \mnras, 444, 3592

\bibitem[{{Doyle} {et~al}\mbox{.}(2013){Doyle}, {Smalley}, {Maxted},
  {Anderson}, {Cameron}, {Gillon}, {Hellier}, {Pollacco}, {Queloz}, {Triaud},
  \& {West}}]{2013MNRAS.428.3164D}
{Doyle} A.~P. {et~al.}, 2013, \mnras, 428, 3164

\bibitem[{{Dumusque}, {Boisse} \& {Santos}(2014){Dumusque}, {.~Boisse}, \&
  {Santos}}]{2014arXiv1409.3594D}
{Dumusque} X., {.~Boisse} I., {Santos} N.~C., 2014, \aap, accepted, arXiv:1409.3594

\bibitem[{{Enoch}, {Collier Cameron} \& {Horne}(2012){Enoch}, {Collier
  Cameron}, \& {Horne}}]{2012A&A...540A..99E}
{Enoch} B., {Collier Cameron} A., {Horne} K., 2012, \aap, 540, A99

\bibitem[{{Feiden} \& {Chaboyer}(2013)}]{2013ApJ...779..183F}
{Feiden} G.~A., {Chaboyer} B., 2013, \apj, 779, 183

\bibitem[{{Fischer} \& {Valenti}(2005)}]{2005ApJ...622.1102F}
{Fischer} D.~A., {Valenti} J., 2005, \apj, 622, 1102

\bibitem[{{Gillon} {et~al}\mbox{.}(2014){Gillon}, {Anderson},
  {Collier-Cameron}, {Delrez}, {Hellier}, {Jehin}, {Lendl}, {Maxted}, {Pepe},
  {Pollacco}, {Queloz}, {S{\'e}gransan}, {Smith}, {Smalley}, {Southworth},
  {Triaud}, {Udry}, {Van Grootel}, \& {West}}]{2014A&A...562L...3G}
{Gillon} M. {et~al.}, 2014, \aap, 562, L3

\bibitem[{{Gillon} {et~al}\mbox{.}(2011){Gillon}, {Doyle}, {Lendl}, {Maxted},
  {Triaud}, {Anderson}, {Barros}, {Bento}, {Collier-Cameron}, {Enoch}, {Faedi},
  {Hellier}, {Jehin}, {Magain}, {Montalb{\'a}n}, {Pepe}, {Pollacco}, {Queloz},
  {Smalley}, {Segransan}, {Smith}, {Southworth}, {Udry}, {West}, \&
  {Wheatley}}]{2011A&A...533A..88G}
---, 2011, \aap, 533, A88

\bibitem[{{Gonzalez}(1997)}]{1997MNRAS.285..403G}
{Gonzalez} G., 1997, \mnras, 285, 403

\bibitem[{{Gray}(2008)}]{2008oasp.book.....G}
{Gray} D.~F., 2008, {The Observation and Analysis of Stellar Photospheres}.
  {Cambridge University Press}

\bibitem[{{H{\'e}brard} {et~al}\mbox{.}(2013){H{\'e}brard}, {Collier Cameron},
  {Brown}, {D{\'{\i}}az}, {Faedi}, {Smalley}, {Anderson}, {Armstrong},
  {Barros}, {Bento}, {Bouchy}, {Doyle}, {Enoch}, {G{\'o}mez Maqueo Chew},
  {H{\'e}brard}, {Hellier}, {Lendl}, {Lister}, {Maxted}, {McCormac}, {Moutou},
  {Pollacco}, {Queloz}, {Santerne}, {Skillen}, {Southworth}, {Tregloan-Reed},
  {Triaud}, {Udry}, {Vanhuysse}, {Watson}, {West}, \&
  {Wheatley}}]{2013A&A...549A.134H}
{H{\'e}brard} G. {et~al.}, 2013, \aap, 549, A134

\bibitem[{{Hellier} {et~al}\mbox{.}(2014){Hellier}, {Anderson}, {Cameron},
  {Delrez}, {Gillon}, {Jehin}, {Lendl}, {Maxted}, {Pepe}, {Pollacco}, {Queloz},
  {S{\'e}gransan}, {Smalley}, {Smith}, {Southworth}, {Triaud}, {Udry}, \&
  {West}}]{2014MNRAS.440.1982H}
{Hellier} C. {et~al.}, 2014, \mnras, 440, 1982

\bibitem[{{Hellier} {et~al}\mbox{.}(2009){Hellier}, {Anderson}, {Collier
  Cameron}, {Gillon}, {Hebb}, {Maxted}, {Queloz}, {Smalley}, {Triaud}, {West},
  {Wilson}, {Bentley}, {Enoch}, {Horne}, {Irwin}, {Lister}, {Mayor}, {Parley},
  {Pepe}, {Pollacco}, {Segransan}, {Udry}, \& {Wheatley}}]{2009Natur.460.1098H}
---, 2009, \nat, 460, 1098

\bibitem[{{Hirano} {et~al}\mbox{.}(2011){Hirano}, {Suto}, {Winn}, {Taruya},
  {Narita}, {Albrecht}, \& {Sato}}]{2011ApJ...742...69H}
{Hirano} T., {Suto} Y., {Winn} J.~N., {Taruya} A., {Narita} N., {Albrecht} S.,
  {Sato} B., 2011, \apj, 742, 69

\bibitem[{{Hoxie}(1973)}]{1973A&A....26..437H}
{Hoxie} D.~T., 1973, \aap, 26, 437

\bibitem[{{Lanza} \& {Shkolnik}(2014)}]{2014MNRAS.443.1451L}
{Lanza} A.~F., {Shkolnik} E.~L., 2014, \mnras, 443, 1451

\bibitem[{{Lendl} {et~al}\mbox{.}(2012){Lendl}, {Anderson}, {Collier-Cameron},
  {Doyle}, {Gillon}, {Hellier}, {Jehin}, {Lister}, {Maxted}, {Pepe},
  {Pollacco}, {Queloz}, {Smalley}, {S{\'e}gransan}, {Smith}, {Triaud}, {Udry},
  {West}, \& {Wheatley}}]{2012A&A...544A..72L}
{Lendl} M. {et~al.}, 2012, \aap, 544, A72

\bibitem[{{L{\'o}pez-Morales}(2007)}]{2007ApJ...660..732L}
{L{\'o}pez-Morales} M., 2007, \apj, 660, 732

\bibitem[{{Lucy} \& {Sweeney}(1971)}]{1971AJ.....76..544L}
{Lucy} L.~B., {Sweeney} M.~A., 1971, \aj, 76, 544

\bibitem[{{Mandel} \& {Agol}(2002)}]{2002ApJ...580L.171M}
{Mandel} K., {Agol} E., 2002, \apjl, 580, L171

\bibitem[{{Marigo} {et~al}\mbox{.}(2008){Marigo}, {Girardi}, {Bressan},
  {Groenewegen}, {Silva}, \& {Granato}}]{2008A&A...482..883M}
{Marigo} P., {Girardi} L., {Bressan} A., {Groenewegen} M.~A.~T., {Silva} L.,
  {Granato} G.~L., 2008, \aap, 482, 883

\bibitem[{{Maxted} {et~al}\mbox{.}(2011){Maxted}, {Anderson}, {Collier
  Cameron}, {Hellier}, {Queloz}, {Smalley}, {Street}, {Triaud}, {West},
  {Gillon}, {Lister}, {Pepe}, {Pollacco}, {S{\'e}gransan}, {Smith}, \&
  {Udry}}]{2011PASP..123..547M}
{Maxted} P.~F.~L. {et~al.}, 2011, \pasp, 123, 547

\bibitem[{{Maxted}, {Serenelli} \& {Southworth}(2014){Maxted}, {Serenelli}, \&
  {Southworth}}]{bagemass}
{Maxted} P.~F.~L., {Serenelli} A.~M., {Southworth} J., 2014, \aap, submitted

\bibitem[{{McQuillan}, {Mazeh} \& {Aigrain}(2013){McQuillan}, {Mazeh}, \&
  {Aigrain}}]{2013ApJ...775L..11M}
{McQuillan} A., {Mazeh} T., {Aigrain} S., 2013, \apjl, 775, L11

\bibitem[{{Nordstr{\"o}m} {et~al}\mbox{.}(2004){Nordstr{\"o}m}, {Mayor},
  {Andersen}, {Holmberg}, {Pont}, {J{\o}rgensen}, {Olsen}, {Udry}, \&
  {Mowlavi}}]{2004A&A...418..989N}
{Nordstr{\"o}m} B. {et~al.}, 2004, \aap, 418, 989

\bibitem[{{Pollacco} {et~al}\mbox{.}(2006){Pollacco}, {Skillen}, {Cameron},
  {Christian}, {Hellier}, {Irwin}, {Lister}, {Street}, {West}, {Anderson},
  {Clarkson}, {Deeg}, {Enoch}, {Evans}, {Fitzsimmons}, {Haswell}, {Hodgkin},
  {Horne}, {Kane}, {Keenan}, {Maxted}, {Norton}, {Osborne}, {Parley}, {Ryans},
  {Smalley}, {Wheatley}, \& {Wilson}}]{2006PASP..118.1407P}
{Pollacco} D.~L. {et~al.}, 2006, \pasp, 118, 1407

\bibitem[{{Popper}(1997)}]{1997AJ....114.1195P}
{Popper} D.~M., 1997, \aj, 114, 1195

\bibitem[{{Queloz} {et~al}\mbox{.}(2000){Queloz}, {Mayor}, {Weber},
  {Bl{\'e}cha}, {Burnet}, {Confino}, {Naef}, {Pepe}, {Santos}, \&
  {Udry}}]{2000A&A...354...99Q}
{Queloz} D. {et~al.}, 2000, \aap, 354, 99

\bibitem[{{Santos}, {Israelian} \& {Mayor}(2004){Santos}, {Israelian}, \&
  {Mayor}}]{2004A&A...415.1153S}
{Santos} N.~C., {Israelian} G., {Mayor} M., 2004, \aap, 415, 1153

\bibitem[{{Serenelli} {et~al}\mbox{.}(2013){Serenelli}, {Bergemann}, {Ruchti},
  \& {Casagrande}}]{2013MNRAS.429.3645S}
{Serenelli} A.~M., {Bergemann} M., {Ruchti} G., {Casagrande} L., 2013, \mnras,
  429, 3645

\bibitem[{{Southworth}(2011)}]{2011MNRAS.417.2166S}
{Southworth} J., 2011, \mnras, 417, 2166

\bibitem[{{Spada} {et~al}\mbox{.}(2013){Spada}, {Demarque}, {Kim}, \&
  {Sills}}]{2013ApJ...776...87S}
{Spada} F., {Demarque} P., {Kim} Y.-C., {Sills} A., 2013, \apj, 776, 87

\bibitem[{{Tegmark} {et~al}\mbox{.}(2004){Tegmark}, {Strauss}, {Blanton},
  {Abazajian}, {Dodelson}, {Sandvik}, {Wang}, {Weinberg}, {Zehavi}, {Bahcall},
  {Hoyle}, {Schlegel}, {Scoccimarro}, {Vogeley}, {Berlind}, {Budavari},
  {Connolly}, {Eisenstein}, {Finkbeiner}, {Frieman}, {Gunn}, {Hui}, {Jain},
  {Johnston}, {Kent}, {Lin}, {Nakajima}, {Nichol}, {Ostriker}, {Pope},
  {Scranton}, {Seljak}, {Sheth}, {Stebbins}, {Szalay}, {Szapudi}, {Xu},
  {Annis}, {Brinkmann}, {Burles}, {Castander}, {Csabai}, {Loveday}, {Doi},
  {Fukugita}, {Gillespie}, {Hennessy}, {Hogg}, {Ivezi{\'c}}, {Knapp}, {Lamb},
  {Lee}, {Lupton}, {McKay}, {Kunszt}, {Munn}, {O'Connell}, {Peoples}, {Pier},
  {Richmond}, {Rockosi}, {Schneider}, {Stoughton}, {Tucker}, {vanden Berk},
  {Yanny}, \& {York}}]{2004PhRvD..69j3501T}
{Tegmark} M. {et~al.}, 2004, \prd, 69, 103501

\bibitem[{{Teitler} \& {K{\"o}nigl}(2014)}]{2014ApJ...786..139T}
{Teitler} S., {K{\"o}nigl} A., 2014, \apj, 786, 139

\bibitem[{{Triaud} {et~al}\mbox{.}(2010){Triaud}, {Collier Cameron}, {Queloz},
  {Anderson}, {Gillon}, {Hebb}, {Hellier}, {Loeillet}, {Maxted}, {Mayor},
  {Pepe}, {Pollacco}, {S{\'e}gransan}, {Smalley}, {Udry}, {West}, \&
  {Wheatley}}]{2010A&A...524A..25T}
{Triaud} A.~H.~M.~J. {et~al.}, 2010, \aap, 524, A25+

\bibitem[{{Weiss} \& {Schlattl}(2008)}]{2008Ap&SS.316...99W}
{Weiss} A., {Schlattl} H., 2008, \apss, 316, 99

\bibitem[{{Winn} {et~al}\mbox{.}(2010){Winn}, {Fabrycky}, {Albrecht}, \&
  {Johnson}}]{2010ApJ...718L.145W}
{Winn} J.~N., {Fabrycky} D., {Albrecht} S., {Johnson} J.~A., 2010, \apjl, 718,
  L145

\bibitem[{{Zhang} \& {Penev}(2014)}]{2014ApJ...787..131Z}
{Zhang} M., {Penev} K., 2014, \apj, 787, 131

\end{thebibliography}

\begin{landscape}

\begin{table} 
\small
\caption{System parameters} 
\label{tab:params}
\begin{tabular}{lcccccc}
\hline
Parameter (Unit) & WASP-87 & WASP-108 & WASP-109 & WASP-110 & WASP-111 & WASP-112 \\ 
\hline 
\\
\multicolumn{7}{l}{Stellar parameters, including from the spectra:}\\
Constellation	& Centaurus							& Centaurus							& Libra								& Sagittarius						& Capricornus						& Piscis Austrinus					\\
Right Ascension	& $\rm 12^{h} 21^{m} 17\fs92$		& $\rm 13^{h} 03^{m} 18\fs73$		& $\rm 15^{h} 28^{m} 13\fs23$		& $\rm 20^{h} 23^{m} 29\fs55$		& $\rm 21^{h} 55^{m} 04\fs23$		& $\rm 22^{h} 37^{m} 57\fs43$		\\
Declination		& $\rm -52\degr 50\arcmin 27\fs0$	& $\rm -49\degr 38\arcmin 22\fs8$	& $\rm -16\degr 24\arcmin 38\fs8$	& $\rm -44\degr 03\arcmin 30\fs3$	& $\rm -22\degr 36\arcmin 45\fs2$	& $\rm -35\degr 09\arcmin 13\fs9$	\\
$V_{\rm mag}$	& 10.7								& 11.2								& 11.4								& 12.3								& 10.3								& 13.3								\\
$K_{\rm mag}$				& 9.6								& 9.8								& 10.2								& 10.7								& 9.0								& 11.9								\\
Spectral type	& F5								& F9								& F4								& G9								& F5								& G6								\\
\teff\ (K)		& $6450 \pm 120$					& $6000 \pm 140$					& $6520 \pm 140$					& $5400 \pm 140$					& $6400 \pm 150$					& $5610 \pm 130$					\\
\logg\ (cgs)	& $4.32 \pm 0.21$					& $4.15 \pm 0.20$					& $4.3 \pm 0.2$						& $4.1 \pm 0.2$						& $4.0 \pm 0.2$						& $4.3 \pm 0.2$						\\
\mictrb\ (\kms)	& $1.34 \pm 0.13$					& $1.2 \pm 0.1$						& $1.6 \pm 0.1$						& $0.9 \pm 0.1$						& $1.5 \pm 0.1$						& $1.0 \pm 0.1$						\\
$v_{\rm mac}$ (\kms) & $5.9 \pm 0.6$				& $4.4 \pm 0.6$						& $6.5 \pm 0.6$						& $3.3 \pm 0.6$						& $6.3 \pm 0.6$						& $3.2 \pm 0.6$						\\
\vsini\ (\kms)	& $9.6 \pm 0.7$						& $4.7 \pm 0.8$						& $15.4 \pm 1.0$					& $0.2 \pm 0.6$						& $11.2 \pm 0.8$					& $2.0 \pm 1.4$						\\
\feh			& $-0.41 \pm 0.10$					& $+0.05 \pm 0.11$					& $-0.22 \pm 0.08$					& $-0.06 \pm 0.10$					& $+0.08 \pm 0.08$					& $-0.64 \pm 0.15$					\\
$\log A$(Li)	& $2.17 \pm 0.08$					& $2.70 \pm 0.12$					& $< 1.1$							& $< 0.6$							& $2.27 \pm 0.11$					& $< 0.7$							\\
Age$_{\rm evol}$ (Gyr)		& $3.8 \pm 0.8$				& $4.6 \pm 1.9$						& $2.6 \pm 0.9$						& $8.6 \pm 3.5$						& $2.6 \pm 0.6$						& $10.6 \pm 3.0$					\\
Distance (pc)	& $240 \pm 20$						& $220 \pm 15$						& $330 \pm 30$						& $320 \pm 30$						& $210 \pm 20$						& $450 \pm 30$						\\
\\
\multicolumn{7}{l}{Parameters from the MCMC analyses:}\\
$P$ (d) & $1.6827950 \pm 0.0000019$ & $2.6755463 \pm 0.0000021$ & $3.3190233 \pm 0.0000042$ & $3.7783977 \pm 0.0000031$ & $2.3109650 \pm 0.0000024$ & $3.0353992 \pm 0.0000038$ \\
$T_{\rm c}$ (HJD$-${\footnotesize 2450000}) & $6361.84030 \pm 0.00021$ & $6413.79019 \pm 0.00015$ & $6361.19263 \pm 0.00023$ & $6502.72413 \pm 0.00016$ & $6275.75127 \pm 0.00040$ & $6505.48734 \pm 0.00013$\\
$T_{\rm 14}$ (d) & $0.12436 \pm 0.00071$ & $0.13216 \pm 0.00052$ & $0.11856 \pm 0.00098$ & $0.11658 \pm 0.00084$ & $0.1373 \pm 0.0017$ & $0.11987 \pm 0.00060$ \\
$T_{\rm 12}=T_{\rm 34}$ (d) & $0.01516 \pm 0.00087$ & $0.01337 \pm 0.00054$ & $0.0237 \pm 0.0010$ & $0.01685 \pm 0.00095$ & $0.0174 \pm 0.0017$ & $0.01642 \pm 0.00069$\\
$a$/\rstar & $3.894 \pm 0.095$ & $7.05 \pm 0.13$ & $7.40 \pm 0.13$ & $11.15 \pm 0.27$ & $4.57 \pm 0.20$ & $8.19 \pm 0.15$ \\
$R_{\rm P}^{2}$/R$_{*}^{2}$ & $0.00765 \pm 0.00013$ & $0.01181 \pm 0.00015$ & $0.01213 \pm 0.00019$ & $0.02085 \pm 0.00031$ & $0.00644 \pm 0.00019$ & $0.01493 \pm 0.00017$\\
$b$ & $0.604 \pm 0.028$ & $0.19 \pm 0.10$ & $0.737 \pm 0.011$ & $0.376 \pm 0.059$ & $0.665 \pm 0.037$ & $0.475 \pm 0.031$\\
$i_{\rm P}$ ($^\circ$) \smallskip & $81.07 \pm 0.63$ & $88.49 \pm 0.84$ & $84.28 \pm 0.19$ & $88.06 \pm 0.35$ & $81.61 \pm 0.82$ & $86.68 \pm 0.28$ \\
$K_{\rm 1}$ (m s$^{-1}$) & $325 \pm 14$ & $117.8 \pm 3.5$ & $109 \pm 15$ & $71.8 \pm 8.2$ & $212 \pm 14$ & $142 \pm 17$ \\
$\gamma$ (m s$^{-1}$) & $-14\,184.5 \pm 7.9$ & $47\,073.925 \pm 2.5$ & $-16\,643 \pm 11$ & $-34\,821.9 \pm 6.9$ & $-19\,800 \pm 11$ & $-19\,587 \pm 11$\\
$e$ & 0 (adopted; $<$ 0.099 at 2\,$\sigma$) & 0 (adopted; $<$ 0.075 at 2\,$\sigma$) & 0 (adopted; $<$ 0.32 at 2\,$\sigma$) & 0 (adopted; $<$ 0.61 at 2\,$\sigma$) & 0 (adopted; $<$ 0.10 at 2\,$\sigma$) & 0 (adopted; $<$ 0.24 at 2\,$\sigma$) \\
$v \sin i_{\rm P}$ (km s$^{-1}$) & --- & --- & --- & --- & $11.12 \pm 0.77$ & ---\\
$\lambda$ ($^\circ$) \smallskip & --- & --- & --- & --- & $-5 \pm 16$& ---\\
$M_{\rm *}$ ($M_{\rm \odot}$) & $1.204 \pm 0.093$ & $1.167 \pm 0.092$ & $1.203 \pm 0.090$ & $0.892 \pm 0.072$ & $1.50 \pm 0.11$ & $0.807 \pm 0.073$\\
$R_{\rm *}$ ($R_{\rm \odot}$) & $1.627 \pm 0.062$ & $1.215 \pm 0.040$ & $1.346 \pm 0.044$ & $0.881 \pm 0.035$ & $1.85 \pm 0.10$ & $1.002 \pm 0.037$\\
$\log g_{*}$ (cgs) & $4.096 \pm 0.023$ & $4.336 \pm 0.018$ & $4.260 \pm 0.018$ & $4.498 \pm 0.022$ & $4.081 \pm 0.036$ & $4.342 \pm 0.018$\\
$\rho_{\rm *}$ ($\rho_{\rm \odot}$) & $0.280 \pm 0.021$ & $0.655 \pm 0.034$ & $0.493 \pm 0.027$ & $1.301 \pm 0.095$ & $0.238 \pm 0.031$ & $0.801 \pm 0.043$\\
$T_{\rm eff}$ (K) & $6480 \pm 110$ & $5960 \pm 120$ & $6520 \pm 140$ & $5360 \pm 130$ & $6470 \pm 120$ & $5650 \pm 110$ \\
{[Fe/H]} \smallskip & $-0.41 \pm 0.10$ & $0.05 \pm 0.11$ & $-0.220 \pm 0.080$ & $-0.06 \pm 0.10$ & $0.081 \pm 0.080$ & $-0.64 \pm 0.15$\\
$M_{\rm P}$ ($M_{\rm Jup}$) & $2.18 \pm 0.15$ & $0.892 \pm 0.055$ & $0.91 \pm 0.13$ & $0.510 \pm 0.064$ & $1.83 \pm 0.15$ & $0.88 \pm 0.12$\\
$R_{\rm P}$ ($R_{\rm Jup}$) & $1.385 \pm 0.060$ & $1.284 \pm 0.047$ & $1.443 \pm 0.053$ & $1.238 \pm 0.056$ & $1.442 \pm 0.094$ & $1.191 \pm 0.049$\\
$\log g_{\rm P}$ (cgs) & $3.416 \pm 0.032$ & $3.093 \pm 0.023$ & $3.000 \pm 0.063$ & $2.881 \pm 0.057$ & $3.305 \pm 0.054$ & $3.153 \pm 0.054$\\
$\rho_{\rm P}$ ($\rho_{\rm J}$) & $0.823 \pm 0.088$ & $0.422 \pm 0.033$ & $0.303 \pm 0.048$ & $0.268 \pm 0.042$ & $0.61 \pm 0.11$ & $0.521 \pm 0.073$\\
$a$ (AU)  & $0.02946 \pm 0.00075$ & $0.0397 \pm 0.0010$ & $0.0463 \pm 0.0011$ & $0.0457 \pm 0.0012$ & $0.03914 \pm 0.00098$ & $0.0382 \pm 0.0011$\\
$T_{\rm P}$ (K) & $2322 \pm 50$ & $1590 \pm 36$ & $1695 \pm 40$ & $1134 \pm 33$ & $2140 \pm 62$ & $1395 \pm 33$\\
\\ 
\hline 
\end{tabular}
\newline Iron abundances are relative to the solar values obtained by \cite{2009ARA&A..47..481A}. 
$v_{\rm mac}$ values obtained using the calibration of \citet{2014MNRAS.444.3592D}. 
Spectral Type estimated from \teff\ using the table in \cite{2008oasp.book.....G}. 
\end{table} 

\label{lastpage}

\end{landscape}

\end{document}